\documentclass[12pt, reqno]{amsart}
\usepackage{amssymb}
\usepackage{mathrsfs}
\usepackage{mathpazo}
\usepackage[colorlinks=true, breaklinks=true]{hyperref}
\usepackage{bm}
\usepackage[latin1]{inputenc}
\usepackage{graphicx}
\newtheorem{theorem}{Theorem}

\newtheorem{corollary}[theorem]{Corollary}

\newtheorem{definition}[theorem]{Definition}

\newtheorem{lemma}[theorem]{Lemma}

\newtheorem{proposition}[theorem]{Proposition}
\newtheorem{remark}[theorem]{Remark}

\newcommand{\bea}{\begin{eqnarray}}
\newcommand{\eq}{\end{eqnarray}}
\newcommand{\eea}{\end{eqnarray}}
\newcommand{\bqn}{\begin{eqnarray*}}
\newcommand{\beaa}{\begin{eqnarray*}}
\newcommand{\eqn}{\end{eqnarray*}}
\newcommand{\eeaa}{\end{eqnarray*}}
\newcommand{\bpr}{\begin{proposition}}
\newcommand{\epr}{\end{proposition}}

\newcommand{\cal}{\mathcal}

\begin{document}
\title{Large deviation principle for Volterra type fractional stochastic volatility models}
\author{Archil Gulisashvili}\let\thefootnote\relax\footnotetext{Department of Mathematics, Ohio University, Athens OH 45701; e-mail: gulisash@ohio.edu}
\date{}

\begin{abstract}
We study fractional stochastic volatility models in which the volatility process is a positive continuous function $\sigma$ of a continuous Gaussian process $\widehat{B}$. Forde and Zhang established a large deviation principle for the log-price process in such a model under the assumptions that the function $\sigma$ is globally H\"{o}lder-continuous and the process $\widehat{B}$ is fractional Brownian motion. In the present paper, we prove a similar small-noise large deviation principle under weaker restrictions on $\sigma$ and $\widehat{B}$. We assume that
$\sigma$ satisfies a mild local regularity condition, while the process $\widehat{B}$ is a Volterra type Gaussian process. Under an additional assumption of the self-similarity of the process $\widehat{B}$, we derive a large deviation principle in the small-time regime. As an application, we obtain asymptotic formulas for binary options, call and put pricing functions, and the implied volatility in certain mixed regimes.

\end{abstract}

\maketitle

\noindent \textbf{AMS 2010 Classification}: 60F10, 60G15, 60G18, 60G22, 41A60, 91G20.\vspace{0.2in%
}

\noindent \textbf{Keywords}: large deviations, Volterra type Gaussian processes, fractional stochastic volatility models, self-similarity, implied volatility. \vspace{%
0.2in}

\section{Introduction}\label{S:in}
In this paper, we study stochastic volatility models in which the volatility process is a continuous 
function of a fractional stochastic process. Typical examples of such processes are fractional Brownian motion, the Riemann-Liouville fractional Brownian motion, and the fractional Ornstein-Uhlenbeck process. We establish small-noise and small-time large deviation principles for fractional models, and also characterize leading terms in asymptotic expansions of option pricing functions and the implied volatility in various regimes.

Over the past few years fractional stochastic volatility models have become increasingly popular. 
In fractional models of our interest in the present paper, the asset price process $S$ satisfies the following stochastic differential equation:
\begin{equation}
dS_t=S_t\sigma(\widehat{B}_t)d(\bar{\rho}W_t+\rho B_t),\quad S_0=s_0> 0,\quad 0\le t\le T,
\label{E:DD}
\end{equation}
where $s_0$ is the initial price, while $T> 0$ is the time horizon. The processes $W$ and $B$ in (\ref{E:DD}) are independent standard Brownian motions, and $\rho\in(-1,1)$ is the correlation coefficient. We also use a standard notation $\bar{\rho}=\sqrt{1-\rho^2}$.
It is assumed in (\ref{E:DD}) that $\sigma$ is a continuous function on $\mathbb{R}$, and $\widehat{B}$ is a continuous fractional stochastic process adapted to the filtration generated by the process $B$ (see the definition of $\widehat{B}$ in (\ref{E:ooo})). The equation in (\ref{E:DD}) is considered on a filtered probability space $(\Omega,\mathcal{F},\{\mathcal{F}_t\}_{0\le t\le T},\mathbb{P})$, where $\{\mathcal{F}_t\}_{0\le t\le T}$ is the filtration generated by $W$ and $B$. The process $\sigma(\widehat{B})$ in 
(\ref{E:DD}) describes the stochastic evolution of volatility in the fractional model. 

We will next discuss the most common fractional processes. \\
\\
\it Fractional Brownian motion: \rm Let $H$ be a number with $0< H< 1$. Fractional Brownian motion $B^H_t$, $t\ge 0$, is a centered Gaussian process with the covariance function given by
\begin{equation}
C_H(t,s)=\frac{1}{2}\left(t^{2H}+s^{2H}-|t-s|^{2H}\right),\quad t,s\ge 0.
\label{E:Cov}
\end{equation}
The process $B^H$ was first implicitly considered by Kolmogorov in \cite{Ko}, and was studied by Mandelbrot and van Ness in \cite{MvN}. The constant $H$ is called the Hurst parameter. It is known that the process $B^H$ has stationary increments. If $H=\frac{1}{2}$, then the process is standard Brownian motion. 
Fractional Brownian motion has a Volterra type representation, that is,
\begin{equation}
B^H_t=\int_0^tK_H(t,s)dB_s,\quad t\ge 0.
\label{E:gib}
\end{equation} 
For $\frac{1}{2}< H< 1$, the kernel $K_H$ in formula (\ref{E:gib}) is defined by
\begin{equation}
K_H(t,s)=c_H\left(H-\frac{1}{2}\right)s^{\frac{1}{2}-H}\int_s^tu^{H-\frac{1}{2}}(u-s)^{H-\frac{3}{2}}du\,
\chi_{\{s< t\}},
\label{E:MG1}
\end{equation}
while for $0< H<\frac{1}{2}$, the kernel $K_H$ is as follows:
\begin{align}
&K_H(t,s) \nonumber \\
&=c_H\left[\left(\frac{t}{s}\right)^{H-\frac{1}{2}}(t-s)^{H-\frac{1}{2}}+\left(\frac{1}{2}-H\right)s^{\frac{1}{2}-H}
\int_s^tu^{H-\frac{3}{2}}(u-s)^{H-\frac{1}{2}}du\right]\chi_{\{s< t\}}.
\label{E:MG2}
\end{align}
In (\ref{E:MG1}) and (\ref{E:MG2}), the function $\chi_{\{s< t\}}$ is defined as follows:
$$
\chi_{\{s< t\}}=\begin{cases}
1,&\mbox{if}\,\,0\le s< t<\infty \\
0,&\mbox{otherwise,}
\end{cases}
$$
while the number $c_H> 0$ is a normalizing constant given by
$$
c_H=\sqrt{\frac{2H\Gamma\left(\frac{3}{2}-H\right)}{\Gamma\left(H+\frac{1}{2}\right)\Gamma(2-2H)}}.
$$
It follows from (\ref{E:gib}) that 
\begin{equation}
C_H(t,s)=\int_0^TK_H(t,u)K_H(s,u)du.
\label{E:cov}
\end{equation}
The Volterra type representation in (\ref{E:gib}) is called the 
Mol\v{c}an-Golosov representation of $B^H$ (see \cite{MG}, p. 135). More details and explanations can be found in 
\cite{DU,NVV,EM}. 
\begin{remark}\label{R:addi}
If $H=\frac{1}{2}$, then the process $B^H$ is a standard Brownian motion. More precisely, $B^{\frac{1}{2}}=B$.
In this case, we have $K_{\frac{1}{2}}(t,s)=1$ for all $0\le s\le t\le T$. Note that formula (\ref{E:MG2}) holds for $H=\frac{1}{2}$.
\end{remark}
\it Riemann-Liouville fractional Brownian motion: \rm For $0< H< 1$, the Riemann-Liouville fractional Brownian motion
is defined by
\begin{equation}
R^H_t=\frac{1}{\Gamma(H+\frac{1}{2})}\int_0^t(t-s)^{H-\frac{1}{2}}
dB_s,\quad t\ge 0.
\label{E:o1}
\end{equation}
This stochastic process was introduced by L\'{e}vy in \cite{PL}. The process in (\ref{E:o1}) is simpler than fractional Brownian motion. However, the increments of the Riemann-Liouville fractional Brownian motion lack the stationarity property. More information about the process $R^H$ can be found in \cite{LS,Pi} \\
\\
\it Fractional Ornstein-Uhlenbeck process: \rm For $0< H< 1$ and $a> 0$, the fractional Ornstein-Uhlenbeck process is given by 
\begin{equation}
U_t^H=\int_0^te^{-a(t-s)}dB^H_s,\quad t\ge 0
\label{E:o2}
\end{equation}
(see \cite{CKM,KS}). The stochastic integral appearing in (\ref{E:o2}) can be defined using the integration by parts formula and the stochastic Fubini theorem. This gives 
$$
U_t^H=B^H_t-a\int_0^te^{-a(t-s)}B_s^Hds
$$
(see, e.g., Proposition A.1 in \cite{CKM}). Therefore,
\begin{equation}
U_t^H=\int_0^t\widehat{K}_H(t,s)dB_s,\quad 0\le t\le T,
\label{E:fOU}
\end{equation}
where 
\begin{equation}
\widehat{K}_H(t,s)=K_H(t,s)-a\int_s^te^{-a(t-u)}K_H(u,s)du,\quad 0\le s< t\le T.
\label{E:kOU}
\end{equation}
Recall that we denoted by $K_H$ the kernel associated with fractional Brownian motion 
(see (\ref{E:MG1}) and (\ref{E:MG2})). Formula (\ref{E:fOU}) provides the Volterra type representation of the process $U^H$, while the function $\widehat{K}_H$ in (\ref{E:kOU}) is the Volterra type kernel associated with $U^H$.

Our next goal is to give a short survey of fractional stochastic volatility models. One of the first continuous-time fractional models with stochastic volatility was introduced  
in the paper \cite{CR} of Comte and Renault. In \cite{CR}, $\sigma(x)=e^x$ and $\widehat{B}$ is the fractional Ornstein-Uhlenbeck process with the Hurst parameter $H>\frac{1}{2}$
(the long memory case). The same process $\widehat{B}$ and a more general function $\sigma$ are used in \cite{CV1,CV2}. In the paper \cite{VMO}, the function $\sigma$ is given by $\sigma(x)=e^{\beta+kx}$, 
where $\beta$ and $k$ are constants, while the process $\widehat{B}$ is as follows: $\widehat{B}_t=\frac{1}{\delta}\left(B^H_t-B^H_{t-\delta}\right)$, $t\ge 0$. In the previous equality, the constant $\delta> 0$ is interpreted as the observation time scale, while the process $\widehat{B}$ is fractional noise. 

The paper \cite{ALV} introduces a fractional model, where the function $\sigma$ satisfies certain boundedness and differentiablity conditions, while the process $\widehat{B}$ 
is the Ornstein-Uhlenbeck process driven by the Riemann-Liouville fBm. Uncorrelated Gaussian and Gaussian self-similar models were introduced in \cite{GVZ1} and \cite{GVZ2}, respectively. In such models
$\sigma(x)=|x|$, while $\widehat{B}$ is a general Gaussian process (\cite{GVZ1}), or a Gaussian self-similar process (\cite{GVZ2}). 

In the groundbreaking paper \cite{GJR}, Gatheral, Jaisson, and Rosenbaum analyzed high-frequency time series of volatility. They claimed that the volatility behaves like the exponential of fBm with $H$ approximately equal to 0.1. 

In \cite{BFG}, a rough Bergomi model was introduced. A simple case of such a model is the following: $\sigma(x)=e^x$ and $\widehat{B}$ is equal to the Riemann-Liouville fBm with $0< H<\frac{1}{2}$. The other papers, where this model is studied, are \cite{JMM,JPS}. In the version of the rough Bergomi model used in \cite{JPS}, the process $\widehat{B}$ is a non-centered Riemann-Liouville fractional Brownian motion. In the present paper, only centered Volterra type processes $\widehat{B}$ are considered. However, in our opinion, it should not be difficult to obtain similar results for fractional models with non-centered processes $\widehat{B}$. 

In \cite{FZ}, the function $\sigma$ satisfies a global H\"{o}lder condition, while $\widehat{B}$ is fBm with $0<H<\frac{1}{2}$. In \cite{F2}, a rough model, in which the function $\sigma$ satisfies certain smoothness and boundedness conditions, and the process $\widehat{B}$ is the Muravlev representation of fractional Brownian motion, 
is studied. The rough model considered in \cite{GaS1,GaS2} uses the function $\sigma$ that is smooth, bounded, and has bounded derivatives, and a scaled fractional Ornstein-Ulenbeck process as the process $\widehat{B}$. The paper \cite{BFGHS} deals with rough stochastic volatility models, in which the function $\sigma$ is smooth, while the acceptable processes $\widehat{B}$ are certain Volterra type Gaussian processes. In an important paper \cite{BFGMS}, rough paths and regularity structures are used to study fractional models. The function $\sigma$ used in \cite{BFGMS} satisfies certain smoothness conditions, while the process $\widehat{B}$ is the Riemann-Liouville fractional Brownian motion. Furthermore, the paper \cite{BFGMS} deals with more complicated fractional models. 

In \cite{BLP}, interesting fractional stochastic volatility models are introduced and studied. In one of them, the log-volatility is modeled by the Cauchy process, that is, the centered stationary Gaussian process with the autocorrelation function given by
$$
A(t)=\left(1+|t|^{2\alpha+1}\right)^{-\frac{\beta}{2\alpha+1}},\quad t\in\mathbb{R},
$$
where $-\frac{1}{2}<\alpha<\frac{1}{2}$ and $\beta> 0$.
In a certain sense, the parameter $\alpha$ describes roughness of the volatility, more precisely, the index of roughness of the model is given by $H=2\alpha+1$. In addition, the parameter $\beta$ characterizes the memory properties of the model. It is important to emphasize that roughness and memory are decoupled in the volatility model described above. In more standard stochastic volatility models, which depend on the Hurst parameter $H$, this effect of decoupling is absent. In such models, roughness and memory properties of the volatility are expressed in terms of the same parameter $H$. The authors of \cite{BLP} also study more complicated non-Gaussian volatility models based on the Cauchy process, or on a Brownian semistationary process (more information can be found in \cite{BLP}).

We finish our incomplete overview of fractional stochastic volatility models by giving references to the so-called fractional Heston models. In such models, the variance process is a fractional version of the CIR-process (the square root process). Fractional Heston models go back to Comte, Coutin, and Renault (see \cite{CCR}). The variance process in \cite{CCR} is the Riemann-Liouville fractional integral operator applied to the CIR-process. A similar variance process is used in \cite{GJRS}, while in \cite{ER1,ER2}, the Riemann-Liouville fractional integral operator is used to modify the stochastic differential equation for the variance in the Heston model. In \cite{JLP}, the Volterra Heston model is introduced and studied. In this model, the Riemann-Liouville kernel is replaced by a more general Volterra type kernel.

We will next comment on what has been accomplished in the present paper. The main results obtained in it
(Theorems \ref{T:1} and \ref{T:smt} below) generalize small-noise and small-time large deviation principles for fractional models, established in the paper \cite{FZ} of Forde and Zhang. Theorem \ref{T:1} contains a small-noise 
large deviation principle for the log-price process $X=\log S$, while Theorem \ref{T:smt} deals with a small-time version. The restrictions imposed on $\sigma$ and $\widehat{B}$ in the present paper are rather mild in comparison with those used in \cite{FZ}. We assume that the function $\sigma$ is locally $\omega$-continuous, where $\omega$ is a given modulus of continuity. We also assume that the process $\widehat{B}$ is a Volterra type Gaussian process (see Definition \ref{D:Volt} in Section \ref{S:cvg}). In \cite{FZ}, the function $\sigma$ satisfies a global H\"{o}lder condition, while the process $\widehat{B}$ is fractional Brownian motion. Large deviation principles, similar to those in \cite{FZ}, were obtained in \cite{BFGMS} for a special class of rough volatility models. In \cite{JPS}, a large deviation principle for the rough Bergomi model was established.

Asymptotic analysis of fractional stochastic volatility models has become a popular field of research in financial mathematics. A significant part of such an analysis is the study of a small maturity behavior of option pricing functions and the implied volatility. 
In \cite{FZ}, Forde and Zhang obtained asymptotic formulas in a mixed small-maturity small-log-moneyness regime for call and put pricing functions and the implied volatility. Using the large deviation principles in Theorems \ref{T:1} and \ref{T:smt}, we derive similar asymptotic formulas in a more general setting (see Section \ref{S:sta}). A great deal of information about the small-maturity asymptotics of various quantities, arising in the theory of fractional models, can be extracted from the papers mentioned above. 

The structure of this paper is as follows. In Section \ref{S:cvg}, we study Gaussian processes
which admit a Volterra type representation with the kernel satisfying the H\"{o}lder condition in $L^2$. We call such processes Volterra type Gaussian processes. Our definition is based on the definition of a Volterra type process in \cite{H,Hu}. Note that an extra restriction is imposed on a Volterra type process in \cite{H,Hu} (see condition (c) in Remark \ref{R:mofus} below). This restriction is not used in the present paper. We also show in Section \ref{S:cvg} that fractional Brownian motion, the Riemann-Liouville fractional Brownian motion, and the fractional Ornstein-Uhlenbeck process are Volterra type processes (see Lemma \ref{L:provide}). In Section \ref{S:Gf}, we discuss fractional stochastic volatility models, and comment on the martingality of the asset price process $S$. In Section \ref{S:MR}, we formulate our main results (Theorems \ref{T:1} and \ref{T:smt}). Theorem \ref{T:1} contains a small-noise large deviation principle for a scaled version of the Volterra type stochastic volatility model. This theorem generalizes a corresponding theorem in \cite{FZ} (see (4.16) and the proof of Theorem 4.8 in Appendix B of \cite{FZ}). However, no additional scaling is used in \cite{FZ}. Although the structure of the proof of Theorem \ref{T:1} is essentially the same as that of the corresponding assertion in \cite{FZ}, various difficulties arise in the case considered in the present paper. The first difficulty is the loss of the stationarity of increments property for the process $\widehat{B}$ (fractional Brownian motion possesses this property, while more general Volterra type Gaussian processes used in this paper do not). One more difficulty is that since our restriction on the function $\sigma$ is local and rather mild, a more refined choice of stopping times is needed in the proof of Theorem \ref{T:1}. In Section \ref{S:MR}, we also derive a small-time large deviation principle using Theorem \ref{T:1}. We use an additional assumption that the process $\widehat{B}$ is self-similar in this derivation (see Theorem \ref{T:smt}). Section \ref{S:first} deals with the following question: Does the large deviation principle remain the same if we remove the drift term in the stochastic differential equation for the log-price? We show in Section \ref{S:first} that the answer to the previous question is affirmative. Section \ref{S:ecp} contains the proof of the small-noise large deviation principle in the case where $T=1$ (the general case is discussed after the formulation of Theorem \ref{T:1} in Section \ref{S:MR}). Finally, Section \ref{S:sta} provides asymptotic formulas for binary options, call and put pricing functions, and the implied volatility in a fractional Volterra type stochastic volatility model.

\section{Volterra type Gaussian processes}\label{S:cvg} 
\vspace{0.1in}
Fix the time horizon $T> 0$, and suppose $K$ is a Lebesgue measurable function on $[0,T]^2$ such that
$$
\int_0^T\int_0^TK(t,s)^2dtds<\infty
$$
(a square integrable kernel). For such a kernel, the linear operator ${\cal K}:L^2[0,T]\mapsto L^2[0,T]$ 
defined by 
$
{\cal K}h(t)=\int_0^TK(t,s)h(s)ds,
$
is compact. The operator ${\cal K}$ is called a Hilbert-Schmidt integral operator.

It will be assumed throughout the paper that
$$
\sup_{t\in[0,T]}\int_0^T|K(t,s)|^2ds<\infty.
$$
Set
\begin{equation}
M(h)=\sup_{\{t_1,t_2\in[0,T]:|t_1-t_2|\le h\}}\int_0^T|K(t_1,s)-K(t_2,s)|^2ds,\quad 0\le h\le T.
\label{E:mofu}
\end{equation}

Denote by $C[0,T]$ the space of continuous functions on the interval $[0,T]$. The norm of a function $f\in C[0,T]$ will be denoted as follows: $||f||_C=\sup_{t\in[0,T]}|f(t)|$. The next assertion is well-known.
\begin{lemma}\label{L:hra}
If $\displaystyle{\lim_{h\downarrow 0}M(h)=0}$, then ${\cal K}$ is a compact linear operator from $L^2[0,T]$ into $C[0,T]$.
\end{lemma}

Let $0<\alpha\le 1$. A function $f\in C[0,T]$ is called H\"{o}lder continuous with exponent $\alpha$ if there exists
$c> 0$ such that 
$
|f(t)-f(s)|\le c|t-s|^{\alpha}
$
for all $t,s\in[0,T]$. The space of all functions, which are H\"{o}lder continuous with exponent $\alpha$, is denoted by $C^{\alpha}[0,T]$. This space, equipped with the norm $||\cdot||_{\alpha}$ defined by
$$
||f||_{\alpha}=||f||_C+\sup_{t,s\in[0,T],t\neq s}\frac{f(t)-f(s)}{|t-s|^{\alpha}},
$$
is a Banach space. It is not hard to prove that if for some $\alpha\in(0,2]$, $M(h)=O(h^{\alpha})$ as $h\downarrow 0$, then the operator ${\cal K}$ is continuous from $L^2[0,T]$ into $C^{\frac{\alpha}{2}}[0,T]$. If for some $\alpha\in(0,2]$, $M(h)=o(h^{\alpha})$ as $h\downarrow 0$, then ${\cal K}:L^2[0,T]\mapsto C^{\frac{\alpha}{2}}[0,T]$ is a compact linear operator. We will not use the previous statement in the present paper.

Suppose $\widehat{B}$ is a centered Gaussian process of the following form:
\begin{equation}
\widehat{B}_t=\int_0^TK(t,s)dB_s,\quad 0\le t\le T,
\label{E:ooo}
\end{equation}
where $K$ is a square integrable kernel, and $B$ is the standard Brownian motion appearing in (\ref{E:DD}).
Denote by $\{\widetilde{\mathcal{F}}_t\}_{0\le t\le T}$ the augmentation of the filtration generated by the process $B$ (see \cite{KaS}, Definition 7.2). The filtration $\{\widetilde{\mathcal{F}}_t\}$ is right-continuous (\cite{KaS}, Corollary 7.8), and the process
$(B_t,\{\widetilde{\mathcal{F}}_t\})$ is still a Brownian motion (\cite{KaS}, Theorem 7.9). 

The covariance function of the process $\widehat{B}$ is given by
$
C(t,s)=\int_0^TK(t,u)K(s,u)du,
$
for $t,s\in[0,T]$.
\begin{definition}\label{D:Volt}
We call the process in (\ref{E:ooo}) a Volterra type Gaussian process if the following conditions hold for the kernel $K$: \\
(a)\,\,$K(0,s)=0$ for all $0\le s\le T$, and $K(t,s)=0$ for all $0\le t< s\le T$.
\\
(b)\,\,There exist constants $c> 0$ and $\alpha> 0$ such that $M(h)\le ch^{\alpha}$ 
for all $h\in[0,T]$ (see (\ref{E:mofu}) for the definition of $M$).
\end{definition}
\begin{remark}\label{R:mofus}
Condition (a) is a typical Volterra type condition for the kernel. Condition (b) was included in the definitions of a Volterra type Gaussian process in \cite{H,Hu} (see Definition 5 in \cite{H} and Definition 5.4 in \cite{Hu}). Note that the definitions in \cite{H,Hu} also contain the following restriction on the kernel $K$: (c)\,\,The operator $\mathcal{K}:L^2[0,T]\mapsto L^2[0,T]$ is injective. Condition (c) will not be used in the present paper.
\end{remark}
\begin{remark}\label{R:vtp}
There is a considerable literature on Volterra type stochastic processes. We will only mention several papers. In \cite{JLP}, affine Volterra processes are introduced and studied. Besides, stochastic convolutions of the following form are considered: $t\mapsto\int_0^tK(t-s)dM_s$, where $M$ is a local martingale, while the kernel $K$ satisfies condition (b) in Definition \ref{D:Volt} (see (2.3) and (2.5) in \cite{JLP}). In \cite{MN}, Volterra type processes with respect to semimartingales are studied under the condition that the kernel $K$ is a function of smooth variation (see \cite{MN}, Definition 1.3). Volterra processes in \cite{D} have the following representation:   $t\mapsto\int_0^tK(t,s)u_sdB_s$, where $u$ is an auxiliary stochastic process. Applications of Volterra type processes to finance are discussed in \cite{ER,JLP}.
\end{remark}

A Volterra type Gaussian process in Definition \ref{D:Volt} has a $\beta$-H\"{o}lder continuous modification for $0<\beta<\frac{\alpha}{2}$. Moreover, such a process is $\widetilde{\mathcal{F}}_t$-adapted (see \cite{H,Hu}). In the sequel, we will always assume that the Volterra type process has $\beta$-H\"{o}lder continuous paths a.s for all $\beta$ with $0<\beta<\frac{\alpha}{2}$. 

In the rest of the present section, we discuss various examples of Volterra type processes. Let us first consider standard Brownian motion $B$ and the Ornstein-Uhlenbeck process $Z_t^{(a)}=\int_0^te^{-a(t-s)}dB_s$, $0\le t\le T$, $a> 0$. The Volterra type kernels associated with these processes are given by $K(t,s)=\chi_{\{s< t\}}$ and $K^{(a)}(t,s)=e^{-a(t-s)}\chi_{\{s< t\}}$,
respectively. It is easy to see that the processes $B$ and $Z_t^{(a)}$ satisfy the conditions in Definition \ref{D:Volt}, and one can take $\alpha=1$ in part (b) of that definition. Hence, Brownian motion and the Ornstein-Uhlenbeck process are Volterra type processes. The previous fact was established in \cite{H,Hu}.
\begin{remark}\label{R:riu}
Standard Brownian motion satisfies condition (c) in Remark \ref{R:mofus} (see \cite{H,Hu}). Indeed, suppose 
a function $f\in L^2[0,T]$ is such that $\int_0^tf(s)ds=0$ for all $t\in[0,T]$. Then, differentiating the previous equality, we obtain $f(t)=0$ a.e. The Ornstein-Uhlenbeck kernel also satisfies condition (c). This can be shown as follows: Suppose $f\in L^2[0,T]$ satisfies $\int_0^te^{-a(t-s)}f(s)ds=0$ for all $t\in[0,T]$. Then, differentiating the previous equality, we obtain $f(t)-a\int_0^te^{-a(t-s)}f(s)ds=0$ a.s. Finally, it follows from the previous two equalities that $f(t)=0$ a.e.
\end{remark}
\begin{remark}\label{R:discr}
It is stated in \cite{H,Hu} that the condition $K(t,s)> 0$ for Lebesgue almost all $0\le s< t\le T$ implies condition (c) in Remark \ref{R:mofus}. In our opinion, the proof of the previous statement given on p. 6 in \cite{H} contains an error. More precisely, there should be the plus sign instead of the minus sign in the formula obtained by differentiating the last displayed formula on p. 6 of \cite{H}. We are not sure whether the proof included in \cite{H} and \cite{Hu} can be corrected.
\end{remark}

Classical fractional processes are of Volterra type. The previous assertion was stated in \cite{H,Hu}. However, no proof of the validity of condition (b) for those processes was included in \cite{H,Hu}, and no information was provided about an acceptable value of the constant $\alpha$. We will next discuss the above-mentioned statement. To prove that the Riemann-Liouville fractional Brownian motion is a Volterra type process is quite straightforward. For fractional Brownian motion, a simple short proof of condition (b) in Definition \ref{D:Volt} with $\alpha=2H$ is given in Section 4 of \cite{Z}. We include the above-mentioned proof from \cite{Z} in the proof of Lemma \ref{L:provide} below for the sake of completeness. It will also be shown in the next lemma that the same statement is valid for the fractional Ornstein-Uhlenbeck process. 
\begin{lemma}\label{L:provide}
The Riemann-Liouville fractional Brownian motion $R^H$, the fractional Brownian motion $B^H$, and the fractional Ornstein-Uhlenbeck process
$U^H$ are Volterra type processes. For all those processes, the constant $\alpha$ in Definition \ref{D:Volt} is given by
$\alpha=2H$. 
\end{lemma} 

\it Proof. \rm We will first prove Lemma \ref{L:provide} for $R^H$. Recall that the kernel $\widetilde{K}_H$ for the process $R^H$ is defined by
$$
\widetilde{K}_H(t,s)=\frac{1}{\Gamma\left(H+\frac{1}{2}\right)}(t-s)^{H-\frac{1}{2}}\chi_{\{s< t\}}. 
$$
Therefore, condition (a) in Definition \ref{D:Volt} is satisfied. It will be shown next that condition (b) also holds. Let us assume that $t+h\le T$. Then we have
\begin{align}
&D_H(h):=\int_0^T\left(\widetilde{K}_H(t+h,s)-\widetilde{K}_H(t,s)\right)^2ds \nonumber \\
&=\frac{1}{\Gamma\left(H+\frac{1}{2}\right)^2}\left[
\int_0^t\left((t+h-s)^{H-\frac{1}{2}}-(t-s)^{H-\frac{1}{2}}\right)^2ds
+\int_t^{t+h}(t+h-s)^{2H-1}ds\right] \nonumber \\
&\le\frac{1}{\Gamma\left(H+\frac{1}{2}\right)^2}\left[
\int_0^T\left((u+h)^{H-\frac{1}{2}}-u^{H-\frac{1}{2}}\right)^2du
+\frac{1}{2H}h^{2H}\right] \nonumber \\
&\le\frac{h^{2H}}{\Gamma\left(H+\frac{1}{2}\right)^2}\left[
\int_0^\infty\left((v+1)^{H-\frac{1}{2}}-v^{H-\frac{1}{2}}\right)^2dv
+\frac{1}{2H}\right].
\label{E:miu}
\end{align}

It is not hard to prove, using the mean value theorem, that the last integral in (\ref{E:miu}) is finite. This shows that the process $R^H$ satisfies condition (b) in Definition \ref{D:Volt} with $\alpha=2H$.

Our next goal is to prove Lemma \ref{L:provide} for $B^H$. The following proof is contained in \cite{Z}. The kernel $K_H$ associated with $B^H$is given in (\ref{E:MG1}) and (\ref{E:MG2}).
Since the covariance function of $B^H$ satisfies (\ref{E:Cov}) and (\ref{E:cov}), we have
\begin{align*}
&\int_0^T(K_H(t,s)-K_H(t^{\prime},s))^2ds=C_H(t,t)-2C(t,t^{\prime})+C_H(t^{\prime},t^{\prime}) \\
&=t^{2H}-\left(t^{2H}+(t^{\prime})^{2H}-|t-t^{\prime}|^{2H}\right)+(t^{\prime})^{2H}=|t-t^{\prime}|^{2H},
\end{align*}
for all $t,t^{\prime}\in[0,T]$. This establishes Lemma \ref{L:provide} for $B^H$.

Finally, we turn our attention to the process $U^H$. It is not hard to see, using formula (\ref{E:kOU}) and 
Lemma \ref{L:provide} for $B^H$, that in order to prove the estimate in condition (b) in Definition \ref{D:Volt} for the kernel
$\widehat{K}_H$, it suffices to show that this condition with $\alpha=2H$ holds for the Volterra type kernel defined by
$$
D_H(t,s)=\int_s^te^{-a(t-u)}K_H(u,s)du\,\chi_{\{s< t\}}.
$$

We have
$$
D_H(t+h,s)-D_H(t,s)=(e^{-ah}-1)\int_s^te^{-a(t-u)}K_H(u,s)du+e^{-ah}\int_t^{t+h}e^{-a(t-u)}K_H(u,s)du.
$$
Therefore, for small $h> 0$,
\begin{align}
&\int_0^T|D_H(t+h,s)-D_H(t,s)|^2ds\le ch^2\int_0^Tds\left(\int_s^te^{-a(t-u)}K_H(u,s)du\right)^2 \nonumber \\
&\quad+c\int_0^Tds\left(\int_t^{t+h}e^{-a(t-u)}K_H(u,s)du\right)^2.
\label{E:jp}
\end{align} 

Suppose $0< H<\frac{1}{2}$. Then, applying the Cauchy-Schwarz inequality to the last integral in (\ref{E:jp}), we obtain
\begin{align}
&\int_0^T|D_H(t+h,s)-D_H(t,s)|^2ds \nonumber \\
&\le ch^2\int_0^T\int_0^TK_H(u,s)^2duds+ch\int_0^T\int_0^TK_H(u,s)^2duds\le ch.
\label{E:sss}
\end{align}
It follows from (\ref{E:sss}) that condition (b) with $\alpha=1$ holds for the kernel $D_H$. Therefore, the same condition holds with 
$\alpha=2H$.

Next, suppose $\frac{1}{2}< H< 1$. Then, using formula (\ref{E:MG1}), we see that
$K_H(t,s)\le c_1s^{\frac{1}{2}-H}$ for all $0\le s\le t\le T$, where $c_1> 0$ is a constant. It follows that
$$
\int_0^Tds\left(\int_t^{t+h}e^{-a(t-u)}K_H(u,s)du\right)^2\le c_2h^2
$$
with $c_2> 0$. The previous estimate and (\ref{E:jp}) show that condition (b) in Definition \ref{D:Volt} holds for the kernel $D_H$
with $\alpha=2H$.

This completes the proof of Lemma \ref{L:provide}.

\section{Fractional stochastic volatility models}\label{S:Gf}
In the rest of the present paper, we assume that the process $\widehat{B}$ in the stochastic model described by (\ref{E:DD}) is a continuous Volterra type Gaussian process (see (\ref{E:ooo}) and Definition \ref{D:Volt}). It is clear that the process $\widehat{B}$ is adapted to the filtration $\left\{{\cal F}_t\right\}_{0\le t\le T}$.

Since the function $\sigma$ is continuous on $\mathbb{R}$ and the process $\widehat{B}$ is continuous, we have
\begin{equation}
\int_0^T\sigma\left(\widehat{B}_s\right)^2ds<\infty
\label{E:squar}
\end{equation}
$\mathbb{P}$-a.s. on $\Omega$. It follows that the equation in (\ref{E:DD}) is a linear stochastic differential equation 
$
dS_t=S_tdM_t
$
with respect to the local martingale 
$$
M_t=\int_0^t\sigma(\widehat{B}_s)d(\bar{\rho}W_s+\rho B_s),\quad 0\le t\le T.
$$
The unique solution to the equation in (\ref{E:DD}) is the Dol\'{e}ans-Dade exponential
\begin{align}
&S_t=s_0\exp\left\{-\frac{1}{2}\int_0^t\sigma^2(\widehat{B}_s)ds+\bar{\rho}\int_0^t\sigma(\widehat{B}_s)dW_s+\rho\int_0^t\sigma(\widehat{B}_s)dB_s\right\},\quad 0\le t\le T.
\label{E:DDE}
\end{align}
Therefore, the log-price
process $X_t=\log S_t$ satisfies
\begin{equation}
X_t=x_0-\frac{1}{2}\int_0^t\sigma^2(\widehat{B}_s)ds+\bar{\rho}\int_0^t\sigma(\widehat{B}_s)dW_s+\rho\int_0^t\sigma(\widehat{B}_s)dB_s,
\label{E:logs}
\end{equation}
where $x_0=\log s_0$. Using (\ref{E:DD}) and (\ref{E:DDE}), we see that the process $S$ is a strictly positive local martingale, and hence a supermartingale. Therefore,
\begin{equation}
\mathbb{E}\left[S_t\right]\le s_0\quad\mbox{for all}\quad t\in[0,T].
\label{E:first}
\end{equation}
The inequality in (\ref{E:first}) means that for every $t\in[0,T]$, the first order moment of the asset price $S_t$ is finite. However, for some Volterra type Gaussian models higher-order moments of the asset price may be infinite (the moment explosion property). For every $t> 0$, we set 
$
p_t^{(m)}=\sup\left\{p> 0:\mathbb{E}\left[S_t^p\right]<\infty\right\}.
$
It follows from (\ref{E:first}) that $p_t^{(m)}\ge 1$. The moment explosion property means that $p_t^{(m)}<\infty$ for all $t\in(0,T]$. For instance, in the uncorrelated Stein-Stein model, where $\sigma(u)=|u|$ and the volatility process $\widehat{B}$ is the Ornstein-Uhlenbeck process (see \cite{SS}), the moments explode (see \cite{GS3}, or  
Theorem 6.17 in \cite{G}). For similar results in the case of the correlated Stein-Stein model with $\sigma(u)=u$, see \cite{DFJV}. In the case of a general uncorrelated Gaussian stochastic volatility model, moment explosions are also present (see Theorem 9 in \cite{GVZ1}). An extreme example of the model explosion feature is the uncorrelated Hull-White model with $\sigma(u)=e^u$ and 
$\widehat{B}$ equal to a standard Brownian motion. Then $p_t^{(m)}=1$ for all $t\in[0,T]$ (see \cite{GS1,GS2}, or \cite{G}, Theorem 6.22).

By taking into account (\ref{E:first}), we see that the asset price process $S$ is a martingale if and only if
$\mathbb{E}\left[S_t\right]=s_0$ for all $t\in[0,T]$. In such a case, $\mathbb{P}$ is a risk-neutral measure.
There are numerous conditions guaranteeing that the Dol\'{e}ans-Dade 
exponential $S$ is a martingale, e.g., Novikov's condition, Kazamaki's condition, Krylov's condition, and Bene\u{s}' 
condition (see, e.g. \cite{KL,KR,LR,MU,RY,R,UZ}).
It is interesting that the asset price process $S$ may be a martingale even for an exponentially growing function $\sigma$ (see \cite{J,L,S}). For instance, it was established in 
\cite{J} that for the Scott model (see \cite{Sc}), where $\sigma(x)=e^x$ and $\widehat{B}$ is the Ornstein-Uhlenbeck process, the process $S$ is a martingale if and only if $-1<\rho\le 0$. Since the Ornstein-Uhlenbeck process with both the initial condition and the long-run mean equal to zero is a Volterra type Gaussian process, the previous statement shows that the asset price process in a Volterra type Gaussian model can be a strict local martingale. The loss of the martingality property in stochastic volatility models was studied in \cite{AP,BKX,Ho,L,LM,S}.

The next lemma shows that if the function $\sigma$ in (\ref{E:DDE}) grows slowly,
then $S$ is a martingale. 
\begin{lemma}\label{L:environ} 
Let $\widehat{B}$ be a continuous
Gaussian process adapted to the filtration $\mathcal{F}_t$, $0\le t\le T$. Denote by $C$ and $m$ the covariance function and the mean function of the process $\widehat{B}$, respectively. Suppose the following nondegeneracy condition holds for the process $\widehat{B}$: $C(t,t)> 0$ for all $t\in(0,T]$. Let $\sigma$ be a positive continuous function on $\mathbb{R}$, satisfying the linear growth condition:
$
\sigma(x)^2\le c_1+c_2x^2,
$
for some $c_1> 0$, $c_2> 0$, and all $x\in\mathbb{R}$.
Then the stochastic exponential $S$ defined by (\ref{E:DDE}) is an $\{%
\mathcal{F}_t\}$-martingale.
\end{lemma}

\it Proof. \rm
Since the process $\widehat{B}$ is continuous, the functions $C$ and $m$ are continuous on $[0,T]^2$ and $[0,T]$, respectively. It suffices to prove that there exists $\delta >0$
such that 
\begin{equation}
L=:\sup_{0<t\leq T}\mathbb{E}[\exp \{\delta\sigma\left(\widehat{B}_{t}\right)^{2}\}]<\infty
\label{E:eq1}
\end{equation}
(see, e.g., \cite{G}, Corollary 2.11). It will be shown that the previous statement
holds provided that 
\begin{equation}
\delta <\left[2c_2\max_{0\leq t\leq T}C(t,t)\right]^{-1}.  \label{E:delta}
\end{equation}

Let $\delta> 0$ be such as in (\ref{E:delta}). Then, we have 
\begin{align}
& L\le e^{c_1\delta}\sup_{0<t\leq T}\mathbb{E}[\exp\{c_2\delta\widehat{B}_{t}^{2}\}] \nonumber \\
&=\frac{e^{c_1\delta}}{\sqrt{2\pi }}\sup_{0<t\leq T}\frac{1}{\sqrt{C(t,t)}}\int_{%
\mathbb{R}}\exp \left\{ c_2\delta y^{2}-\frac{1}{2C(t,t)}(y-m(t))^{2}\right\}dy.
\label{E:shri}
\end{align}%
Set $\alpha (t)=\frac{1}{2C(t,t)}-c_2\delta $ and $\beta (t)=\frac{m(t)}{C(t,t)}
$. Then, transforming the expression on the right-hand side of (\ref{E:shri}), we see that
\begin{align*}
& L\leq\frac{e^{c_1\delta}}{\sqrt{2\pi }}\sup_{0<t\leq T}\frac{1}{\sqrt{C(t,t)}}\exp \left\{ 
\frac{\beta (t)^{2}}{4\alpha (t)}-\frac{m(t)^{2}}{2C(t,t)}\right\} \int_{%
\mathbb{R}}\exp \left\{ -\alpha (t)y^{2}\right\} dy \\
& =\frac{e^{c_1\delta}}{\sqrt{2}}\sup_{0<t\leq T}\frac{1}{\sqrt{C(t,t)\alpha (t)}}\exp \left\{ \frac{%
\beta (t)^{2}}{4\alpha (t)}-\frac{m(t)^{2}}{2C(t,t)}\right\} \\
&=e^{c_1\delta}\sup_{0<t\leq T}\frac{1}{\sqrt{1-2c_2\delta C(t,t)}}\exp \left\{ \frac{%
c_2\delta m(t)^{2}}{1-2c_2\delta C(t,t)}\right\} \\
& \leq \frac{e^{c_1\delta}}{\sqrt{1-2c_2\delta \max_{0\leq t\leq T}C(t,t)}}\exp
\left\{ \frac{c_2\delta m(t)^{2}}{1-2c_2\delta \max_{0\leq t\leq T}C(t,t)}\right\}.
\end{align*}%
Since the mean function $m$ is continuous, we have $|m(t)|\le M$ for all $0\leq
t\leq T$. Finally, using (\ref{E:delta}) and the previous inequality, we obtain (\ref{E:eq1}).

This completes the proof of Lemma \ref{L:environ}.
\begin{remark}\label{R:pozd}
Suppose the stochastic exponential $S$ used in Lemma \ref{L:environ} is a strict local martingale. Since $S$ is a continuous adapted process, it is locally bounded. It follows from Corollary 1.2 in \cite{DeS} that $S$ satisfies the condition of No Free Lunch with Vanishing Risk (see \cite{DeS} for the necessary definitions and more details). 
\end{remark}
\section{Small-noise and small-time large deviation principles}\label{S:MR}
Fix a parameter $H> 0$, and for every $\varepsilon\in(0,1]$, consider the following scaled version of the stochastic differential 
equation in (\ref{E:DD}):
\begin{equation}
dS_t^{\varepsilon,H}=\sqrt{\varepsilon}S_t^{\varepsilon,H}\sigma(\varepsilon^{H}\widehat{B}_t)d(\bar{\rho}W_t+\rho B_t),
\quad S_0^{\varepsilon,H}=s_0.
\label{E:scaled}
\end{equation}
For every $\beta\in(0,1]$, we have
$
\int_0^T\sigma\left(\beta\widehat{B}_s\right)^2ds<\infty
$
$\mathbb{P}$-a.s. on $\Omega$.
Moreover, the process $X_t^{\varepsilon,H}=\log S^{\varepsilon,H}_t$, $0\le t\le T$, with $X_0^{\varepsilon,H}=x_0=\log s_0$ satisfies
\begin{align}
X_t^{\varepsilon,H}&=x_0-\frac{1}{2}\varepsilon\int_0^t\sigma^2(\varepsilon^{H}\widehat{B}_s)ds
+\sqrt{\varepsilon}\bar{\rho}\int_0^t\sigma(\varepsilon^{H}\widehat{B}_s)dW_s 
+\sqrt{\varepsilon}\rho\int_0^t\sigma(\varepsilon^{H}\widehat{B}_s)dB_s.
\label{E:firs}
\end{align}

We will need the following definition.
\begin{definition}\label{D:moc}
Let $\omega$ be an increasing modulus of continuity on $[0,\infty)$, that is, $\omega:\mathbb{R}^{+}\mapsto\mathbb{R}^{+}$ is an increasing function such that $\omega(0)=0$ and 
$\displaystyle{\lim_{s\downarrow 0}\omega(s)=0}$. We call a function $\sigma$ defined on $\mathbb{R}$ locally $\omega$-continuous,
if for every $\delta> 0$ there exists $L(\delta)> 0$ such that 
$$
|\sigma(x)-\sigma(y)|\le L(\delta)\omega(|x-y|)
$$ 
for all $x,y\in[-\delta,\delta]$. 
\end{definition}

A special example of a modulus of continuity is $\omega(s)=s^{\alpha}$ with $\alpha\in(0,1)$. In this case, we get a local $\alpha$-H\"{o}lder condition. If $\alpha=1$, then the condition in Definition \ref{D:moc} is a local Lipschitz condition.
\begin{remark}\label{R:beggo}
With no loss of generality, we may suppose that $\delta\mapsto L(\delta)$ is an even strictly increasing continuous function on $[0,\infty)$ with $L(0)> 0$ and
$
\displaystyle{\lim_{\delta\rightarrow\infty}L(\delta)=\infty}.
$ 
Then, the inverse function $L^{-1}$ is defined and continuous on 
$[L(0),\infty)$. Moreover,
$
\displaystyle{\lim_{\gamma\rightarrow\infty}L^{-1}(\gamma)=\infty}.
$
We will assume throughout the paper that the function $L$ satisfies the conditions formulated above.
\end{remark}

Let $f\in H^1_0[0,T]$, where the symbol $H^1_0[0,T]$ stands for the Cameron-Martin space, consisting of absolutely continuous functions $f$ on $[0,T]$ such that $f(0)=0$ and $\dot{f}\in L^2[0,T]$. The following notation will be used in the sequel:
\begin{equation}
\widehat{f}(s)=\int_0^sK(s,u)\dot{f}(u)du.
\label{E:kernel}
\end{equation}

We will next recall the definition of a rate function on $\mathbb{R}$ (more details can be found in \cite{DZ}). An extended real function $I:\mathbb{R}\mapsto[0,\infty]$ is called a rate function if it is not identically $\infty$ and is lower semi-continuous. The latter property of $I$ means that for every $\delta> 0$, the sub-level set $\Gamma_{\delta}=\left\{x\in\mathbb{R}:I(x)\le\delta\right\}$ is a closed subset of $\mathbb{R}$. If for every $\delta> 0$, the set $\Gamma_{\delta}$ is a compact subset of $\mathbb{R}$, then $I$ is called a good rate function. 

The next assertion is one of the main results of the present paper. It provides a generalization of a large deviation principle established in \cite{FZ}. 
\begin{theorem}\label{T:1}
Suppose $\sigma$ is a positive function on $\mathbb{R}$ that is locally $\omega$-continuous for some modulus of continuity $\omega$.
Let $H> 0$, and let $\widehat{B}$ be a Volterra type Gaussian process. Set
\begin{equation}
I_T(x)=\inf_{f\in H_0^1[0,T]}\left[\frac{\left(x-\rho\int_0^T\sigma(\widehat{f}(s))\dot{f}(s)ds\right)^2}
{2\bar{\rho}^2\int_0^T\sigma(\widehat{f}(s))^2ds}+\frac{1}{2}\int_0^T\dot{f}(s)^2ds\right].
\label{E:vunz}
\end{equation}
Then the function $I_T$ is a good rate function. Moreover, a small-noise large deviation principle with speed $\varepsilon^{-2H}$ and rate function $I_T$ given by (\ref{E:vunz}) holds for the process
$\varepsilon\mapsto\varepsilon^{H-\frac{1}{2}}\left(X_T^{\varepsilon,H}-x_0\right)$, $0<\varepsilon\le 1$, where $X_T^{\varepsilon,H}$ is defined by (\ref{E:firs}). More precisely,
for every Borel measurable subset $A$ of $\mathbb{R}$, the following estimates hold:
\begin{align}
&-\inf_{x\in A^{\circ}}I_T(x)\le\liminf_{\varepsilon\downarrow 0}\varepsilon^{2H}\log\mathbb{P}\left(\varepsilon^{H-\frac{1}{2}}\left(X_T^{\varepsilon,H}-x_0\right)\in A\right) 
\nonumber \\
&\le\limsup_{\varepsilon\downarrow 0}\varepsilon^{2H}\log\mathbb{P}\left(\varepsilon^{H-\frac{1}{2}}\left(X_T^{\varepsilon,H}-x_0\right)\in A\right)
\le-\inf_{x\in\bar{A}}I_T(x).
\label{E:liu}
\end{align}
The symbols $A^{\circ}$ and $\bar{A}$ in the previous estimates stand for the interior and the closure of the set $A$, respectively.
\end{theorem}
\begin{remark}\label{R:oo}
The proof of the fact that the function $I_1$ is a good rate function is given in Section \ref{S:ecp} (see Remark \ref{R:grf2}). For the function $I_T$ with $T\neq 1$, the same statement can be obtained, using the reasoning showing how to prove Theorem \ref{T:1} for $T\neq 1$, assuming that it holds for $T=1$ (see the discussion before Definition \ref{D:ses}).
\end{remark}

In the next assertion, we discuss various properties of the function $I_T$.
\begin{lemma}\label{L:conti}
The rate function $I_T:\mathbb{R}\mapsto[0,\infty)$ defined in (\ref{E:vunz}) is continuous on $\mathbb{R}$.
Moreover, the function $I_T$ is non-increasing on $(-\infty,0]$, non-decreasing on $[0,\infty)$, and $I_T(0)=0$.
\end{lemma}
\begin{remark}
The proof of the monotonicity properties of the rate function $I_T$ given below is a simple modification of the proof of a similar statement in a slightly different setting due to Christian Bayer (private communication, December 2017). We are grateful to Christian for sharing his proof.
\end{remark}

\it Proof of Lemma \ref{L:conti}. \rm 
The fact that $I_T$ is continuous is rather standard (see, e.g., Corollary 4.6 in \cite{FZ}). We include its proof for the sake of completeness. Since $I_T$ is a rate function, it is lower-semicontinuous. On the other hand, the function $I_T$ is equal to the infimum of a family of continuous functions, and hence it is upper-semicontinuous. It follows that $I_T$ is a continuous function. It is also clear that the function $I_T$ is everywhere non-negative. Next, using the function $f=0$, we see that $I_T(0)=0$. It remains to prove the monotonicity statements in Lemma \ref{L:conti}. We will only prove that $I_1$ is a non-decreasing function on $(0,\infty)$. The proof in the remaining cases is similar.

The representation of the function $I_1$ given in formulas (\ref{E:vu}) and (\ref{E:lab}) below will be needed in the proof. We will reason by contradiction. Let $x_1< x_2$, and suppose $I_1(x_2)< I_1(x_1)$.
Fix $\tau> 0$, for which $I_1(x_2)+\tau< I_1(x_1)$. Then, it follows 
from (\ref{E:vu}) and (\ref{E:lab}) that there exist $y\in\mathbb{R}$ and $f\in H_0^1[0,1]$, which depend on $\tau$, and are such that
$\Phi(y,f,\hat{f})=x_2$
and 
\begin{align}
&I_1(x_2)\le\frac{1}{2}y^2+\int_0^1\dot{f}(s)^2ds<\frac{1}{2}z^2+\int_0^1\dot{g}(s)^2ds,
\label{E:ali}
\end{align}
for all $z\in\mathbb{R}$ and $g\in H_0^1[0,1]$, satisfying the condition
$\Phi\left(z,g,\hat{g}\right)=x_1$.

For any $0\le t\le 1$, $x\in\mathbb{R}$ and $f\in H_0^1[0,1]$, set
$$
\Phi_t(x,f,\widehat{f})=\bar{\rho}
\left\{\int_0^t\sigma(\widehat{f}(s))^2ds\right\}^{\frac{1}{2}}x+\rho
\int_0^t\sigma(\widehat{f}(s))\dot{f}(s)ds.
$$
The function $t\mapsto\Phi_t\left(y,f,\hat{f}\right)$ is continuous.
Moreover,
$\Phi_0\left(y,f,\hat{f}\right)=0$ and
$$
\Phi_1\left(y,f,\hat{f}\right)=\Phi(y,f,\hat{f})=x_2.
$$
Next, using the assumption $0< x_1< x_2$ and the intermediate value theorem, we see that
there exists $r\in(0,1)$, depending on $\varepsilon$ and such that 
$\Phi_r(y,f,\hat{f})=x_1$.

Define a function $h$ as follows: $h(s)=f(s)$ for $0\le s\le r$
and $h(s)=f(r)$ for $r\le s\le 1$.
Then $h\in H_0^1[0,1]$, 
$\Phi(y,h,\hat{h})=\Phi_r(y,f,\hat{f})=x_1$, and $\int_0^1\dot{h}(s)^2ds\le\int_0^1\dot{f}(s)^2ds$.
Finally, taking $z=y$ and $g=h$ in (\ref{E:ali}), we arrive at a contradiction. 

This completes the proof of Lemma \ref{L:conti}.

The proof of Theorem \ref{T:1} in the special case where $T=1$ is contained in Section \ref{S:ecp}. 
It resembles the proof of the small-noise analogue of Theorem 4.5 in \cite{FZ}. However, as it has already been mentioned in the introduction, we had to overcome two major difficulties. It is assumed in \cite{FZ} that the function $\sigma$ satisfies the global H\"{o}lder condition, while our restriction on the regularity of $\sigma$ is local and rather weak. Moreover, the process $\widehat{B}$ in \cite{FZ} is fractional Brownain motion, which allows the authors of \cite{FZ} to use the fact that the increments of fractional Brownian motion are stationary. In the present paper, the process $\widehat{B}$ is a general continuous Volterra type Gaussian process, and among such processes, only fractional Brownian motion has stationary increments. 

Our next goal is to explain how to prove Theorem \ref{T:1} in the general case, assuming that it holds for $T=1$.
The Volterra kernel $K$ in Theorem \ref{T:1} is defined on the set $[0,T]^2$. Set
\begin{equation}
\widetilde{K}_T(r,u)=T^{\frac{1}{2}-H}K(Tr,Tu),\quad 0\le r,u\le 1.
\label{E:equa}
\end{equation}
Then $\widetilde{K}_T$ is a Volterra type kernel on $[0,1]^2$.

Consider the process $\varepsilon\mapsto\widetilde{X}_1^{\varepsilon,H}$, $0<\varepsilon\le 1$, associated with the kernel $\widetilde{K}_T$. Then for every $\varepsilon\in(0,1]$, we have 
\begin{equation}
\varepsilon^{H-\frac{1}{2}}X_T^{\frac{\varepsilon}{T},H}=\varepsilon^{H-\frac{1}{2}}\widetilde{X}_1^{\varepsilon,H}
\label{E:3}
\end{equation}
(we leave the previous equality as an exercise for the interested reader). Applying Theorem \ref{T:1} with $T=1$ to the process on the right-hand side of (\ref{E:3}), 
we see that for every Borel set $D\subset\mathbb{R}$,
\begin{align*}
&-\inf_{y\in D^{\circ}}\widehat{I}_T(y)\le\liminf_{\varepsilon\downarrow 0}\varepsilon^{2H}\log\mathbb{P}\left(\varepsilon^{H-\frac{1}{2}}\left(X_T^{\frac{\varepsilon}{T},H}-x_0\right)\in D\right) \\
&\le\limsup_{\varepsilon\downarrow 0}\varepsilon^{2H}\log\mathbb{P}\left(\varepsilon^{H-\frac{1}{2}}\left(X_T^{\frac{\varepsilon}{T},H}-x_0\right)\in D\right)
\le-\inf_{y\in\bar{D}}\widehat{I}_T(y),
\end{align*}
where
\begin{align}
&\widehat{I}_T(y)=\inf_{h\in H_0^1[0,1]}\left[\frac{\left(y-\rho\int_0^1\sigma\left(\int_0^1\widetilde{K}_T(s,u)\dot{h}(u)du\right)\dot{h}(s)ds\right)^2}
{2\bar{\rho}^2\int_0^1\sigma\left(\int_0^1\widetilde{K}_T(s,u)\dot{h}(u)du\right)^2ds}+\frac{1}{2}\int_0^1\dot{h}(s)^2ds\right].
\label{E:dop}
\end{align}

Suppose $A$ is a Borel set in $\mathbb{R}$. Then, replacing $\varepsilon$ by $T\varepsilon$ and $D$ by $T^{H-\frac{1}{2}}A$,
we obtain
\begin{align*}
&-\inf_{x\in A^{\circ}}\left(T^{-2H}\widehat{I}_T(T^{H-\frac{1}{2}}x)\right)\le\liminf_{\varepsilon\downarrow 0}\varepsilon^{2H}\log\mathbb{P}\left(\varepsilon^{H-\frac{1}{2}}\left(X_T^{\varepsilon,H}-x_0\right)\in A\right) \\
&\le\limsup_{\varepsilon\downarrow 0}\varepsilon^{2H}\log\mathbb{P}\left(\varepsilon^{H-\frac{1}{2}}\left(X_T^{\varepsilon,H}-x_0\right)\in A\right)
\le-\inf_{x\in\bar{A}}\left(T^{-2H}\widehat{I}_T(T^{H-\frac{1}{2}}x)\right).
\end{align*}
It remains to show that for every $x\in\mathbb{R}$,
\begin{equation}
T^{-2H}\widehat{I}_T(T^{H-\frac{1}{2}}x)=I_T(x).
\label{E:na1}
\end{equation}

Let $f\in H_0^1[0,T]$, and set $h(u)=T^{H-\frac{1}{2}}f(Tu)$, $0\le u\le 1$. Then $f\leftrightarrow h$ 
is a one-to-one correspondence between $H_0^1[0,T]$ and $H_0^1[0,1]$. Using (\ref{E:equa}) and the previous remark 
in (\ref{E:dop}), and simplifying the resulting expression, we obtain 
\begin{equation}
\widehat{I}_T(y)=\inf_{f\in H_0^1[0,T]}
J(y,f),
\label{E:vvu}
\end{equation} 
where
\begin{align}
&J(y,f) 
=\frac{\left[y-\rho T^{H-\frac{1}{2}}\int_0^T\sigma\left(\int_0^zK(z,y)\dot{f}(y)dy\right)\dot{f}(z)dz\right]^2}
{2\bar{\rho}^2T^{-1}\int_0^T\sigma\left(\int_0^zK(z,y)\dot{f}(y)dy\right)^2dz} 
+\frac{1}{2}T^{2H}\int_0^T\dot{f}(y)^2dy.
\label{E:vuz}
\end{align}
Now, it is clear that (\ref{E:na1}) follows from (\ref{E:vvu}), (\ref{E:vuz}), and (\ref{E:vunz}).

The previous reasoning shows how to derive Theorem \ref{T:1} in the general case from the case where $T=1$.

We will next assume that the process $\widehat{B}$ is self-similar. This additional restriction will allow us to derive a small-time large deviation principle for the process $t\mapsto t^{H-\frac{1}{2}}\left(X_t-x_0\right)$, $0< t\le T$,
from the small-noise large deviation principle established in Theorem \ref{T:1}. 
\begin{definition}\label{D:ses}
For $0< H< 1$, the process $\widehat{B}_t$, $0\le t\le T$, is called $H$-self-similar if for every
$\varepsilon\in(0,1]$,
$\widehat{B}_{\varepsilon t}=\varepsilon^H\widehat{B}_t$, $0\le t\le T$,
in law. 
\end{definition}

It often happens that under certain self-similarity assumptions one can pass from the small-noise LDP to a corresponding small-time LDP.
In our case, if the process $\widehat{B}$ is $H$-self-similar, then it is not hard to see, using (\ref{E:logs}), (\ref{E:firs}), and the $\frac{1}{2}$-self-similarity of Brownian motion, that 
\begin{equation}
X_T^{\varepsilon,H}=X_{\varepsilon T},\quad 0<\varepsilon\le 1.
\label{E:ssp}
\end{equation}
It follows from (\ref{E:ssp}) that the process $\varepsilon\mapsto\varepsilon^{H-\frac{1}{2}}\left(X_T^{\varepsilon,H}-x_0\right)$, $0<\varepsilon\le 1$, in Theorem \ref{T:1} can be replaced by the process $\varepsilon\mapsto\varepsilon^{H-\frac{1}{2}}\left(X_{\varepsilon T}-x_0\right)$, $0<\varepsilon\le 1$. Let us set $t=\varepsilon T$ in the resulting theorem. Then $t\in[0,T]$, and instead of the rate function $I_T$, we obtain a new rate function given by $x\mapsto T^{2H}I_T\left(T^{\frac{1}{2}-H}x\right)$, $x\in\mathbb{R}$. It follows from (\ref{E:na1}) that the new rate function coincides with the function $\widehat{I}_T$ defined in (\ref{E:dop}). Next, summarizing what was said above, we see that
the following small-time large deviation principle holds:
\begin{theorem}\label{T:smt}
Suppose $\sigma$ is a positive function on $\mathbb{R}$ that is locally $\omega$-continuous for some modulus of continuity $\omega$.
Suppose also that $\widehat{B}$ is an $H$-self-similar Volterra type Gaussian process, where $0< H< 1$.
Then a small-time LDP with speed $t^{-2H}$ and good rate function $\widehat{I}_T$ given by (\ref{E:dop}) holds for the process $t\mapsto t^{H-\frac{1}{2}}\left(X_t-x_0\right)$, $0< t\le T$, where $X$ is defined by (\ref{E:logs}). More precisely,
for every Borel measurable set $A\subset\mathbb{R}$, 
\begin{align*}
&-\inf_{x\in A^{\circ}}\widehat{I}_T(x)\le\liminf_{t\downarrow 0}t^{2H}\log\mathbb{P}\left(t^{H-\frac{1}{2}}\left(X_t-x_0\right)\in A\right) \\
&\le\limsup_{t\downarrow 0}t^{2H}\log\mathbb{P}\left(t^{H-\frac{1}{2}}\left(X_t-x_0\right)\in A\right)
\le-\inf_{x\in\bar{A}}\widehat{I}_T(x).
\end{align*}
\end{theorem}
\begin{remark}\label{R:impo}
In this remark, we show that in a sense, the function $\widehat{I}_T$ does not depend on $T$. Let $0< T_1\le T$. Then the kernel $K$, initially defined on $[0,T]^2$, can be restricted to $[0,T_1]^2$, and hence the process $t\mapsto\widehat{B}_t$ can be considered on the interval $[0,T_1]$. We will next show that if the process $t\mapsto\widehat{B}_t$, $0\le t\le T$, is $H$-self-similar, then $\widehat{I}_T(x)=\widehat{I}_{T_1}(x)$ for all $x\in\mathbb{R}$. Indeed, the self-similarity of $\widehat{B}$ implies that $K(t,s)=\varepsilon^{\frac{1}{2}-H}K(\varepsilon t,
\varepsilon s)$ for all $0\le s\le t\le T$ and $0<\varepsilon\le 1$. Therefore, $T_1^{\frac{1}{2}-H}K(T_1r,T_1u)
=T^{\frac{1}{2}-H}K(Tr,Tu)$ for all $0\le u\le r\le 1$, and hence $\widetilde{K}_{T_1}(r,u)=\widetilde{K}_T(r,u)$
for all $0\le u\le r\le 1$, where $\widetilde{K}_T$ is defined in (\ref{E:equa}). Next, we see that
(\ref{E:dop}) implies that $\widehat{I}_T(x)=\widehat{I}_{T_1}(x)$ for all $x\in\mathbb{R}$. 
\end{remark}
\section{The small-noise LDP does not change if the drift term is removed}\label{S:first}
Suppose $\widehat{B}$ is a Gaussian process such as in (\ref{E:ooo}). Suppose also that $\sigma$ is a positive and continuous function on $\mathbb{R}$. It is not hard to prove that there exists a continuous positive even function $\eta$ on $\mathbb{R}$, satisfying the following conditions: $\eta$ is strictly increasing on $[0,\infty)$; $\displaystyle{\lim_{u\rightarrow\infty}\eta(u)=\infty}$; and $\sigma(u)^2\le\eta(u)$ for all $u\in\mathbb{R}$. We will denote by $\eta^{-1}$ the inverse function of the function $\eta(x)$, $x\in[0,\infty)$. The function $\eta^{-1}$ is defined on $[\eta(0),\infty)$ and maps the previous set onto $[0,\infty)$.

Compare the SDE in (\ref{E:firs}) with the following SDE:
$$
d\widehat{X}_t^{\varepsilon,H}=\sqrt{\varepsilon}\sigma(\varepsilon^{H}\widehat{B}_t)d(\bar{\rho}W_t+\rho B_t),\quad
{X}_0^{\varepsilon,H}=x_0.
$$
The solution to the previous equation is given by
$$
\widehat{X}_t^{\varepsilon,H}=x_0+\sqrt{\varepsilon}\bar{\rho}\int_0^t\sigma(\varepsilon^{H}\widehat{B}_s)dW_s+
\sqrt{\varepsilon}\rho\int_0^t\sigma(\varepsilon^{H}\widehat{B}_s)dB_s,
$$
and it follows that
$$
X_T^{\varepsilon,H}-\widehat{X}_T^{\varepsilon,H}=-\frac{1}{2}\varepsilon\int_0^T\sigma^2(\varepsilon^{H}\widehat{B}_s)ds.
$$
Therefore, for every $\delta> 0$,
\begin{align*}
&\mathbb{P}\left(\left|\varepsilon^{H-\frac{1}{2}}X_T^{\varepsilon,H}-\varepsilon^{H-\frac{1}{2}}\widehat{X}_T^{\varepsilon,H}
\right|>\delta\right)=\mathbb{P}\left(\frac{1}{2}\varepsilon^{H+\frac{1}{2}}\int_0^T
\sigma^2(\varepsilon^{H}\widehat{B}_s)ds>\delta\right) \\
&\le\mathbb{P}\left(\eta\left(\varepsilon^{H}\sup_{t\in[0,T]}\left|\widehat{B}_t\right|\right)
>2\delta\varepsilon^{-(H+\frac{1}{2})}T^{-1}\right)  \\
&=\mathbb{P}\left(\sup_{t\in[0,T]}\left|\widehat{B}_t\right|
>\varepsilon^{-H}\eta^{-1}\left(2\delta\varepsilon^{-(H+\frac{1}{2})}T^{-1}\right)\right).
\end{align*}

Using the large deviation principle for the maximum of a Gaussian process (see, e.g., (8.5) in \cite{L2}), we can show that there exist constants $C_1> 0$ and $y_0> 0$ such that
\begin{equation}
\mathbb{P}\left(\sup_{t\in[0,T]}|\widehat{B}_t|> y\right)\le e^{-C_1 y^2}
\label{E:addik}
\end{equation}
for all $y> y_0$. It is not hard to prove that
$$
\lim_{\varepsilon\downarrow 0}\varepsilon^{-H}\eta^{-1}\left(2\delta\varepsilon^{-(H+\frac{1}{2})}T^{-1}\right)
=\infty.
$$
Therefore, there exists $\varepsilon_0> 0$ such that for $\varepsilon<\varepsilon_0$, we have
\begin{align}
&\mathbb{P}\left(\left|\varepsilon^{H-\frac{1}{2}}X_T^{\varepsilon,H}-\varepsilon^{H-\frac{1}{2}}
\widehat{X}_T^{\varepsilon,H}
\right|>\delta\right)\le \exp\left\{-C_1\varepsilon^{-2H}\eta^{-1}\left(2\delta\varepsilon^{-(H+\frac{1}{2})}T^{-1}\right)^2\right\}.
\label{E:fina}
\end{align}
Now, it is easy to see using (\ref{E:fina}) that 
$$
\limsup_{\varepsilon\downarrow 0}\left[\varepsilon^{2H}\log\mathbb{P}\left(\left|\varepsilon^{H-\frac{1}{2}}X_T^{\varepsilon,H}-\varepsilon^{H-\frac{1}{2}}\widehat{X}_T^{\varepsilon,H}
\right|>\delta\right)\right]=-\infty.
$$

It follows that the processes 
$
\varepsilon\mapsto\varepsilon^{H-\frac{1}{2}}\left(X_T^{\varepsilon,H}-x_0\right)$ and
$\varepsilon\mapsto\varepsilon^{H-\frac{1}{2}}\left(\widehat{X}_T^{\varepsilon,H}-x_0\right)$
are exponentially equivalent (see Definition 4.2.10 in \cite{DZ}). Therefore, if the large deviation principle with speed $\varepsilon^{-2H}$ and rate function $I_T$ holds for the latter process, it also holds for the former one
(see Theorem 4.2.13 in \cite{DZ}).
\section{Proof of the small-noise large deviation principle}\label{S:ecp} 
Let us assume that $\rho\neq 0$. The proof for $\rho=0$ uses the same ideas and is much simpler. 
In the present section, we will prove Theorem \ref{T:1} in the case where $T=1$.

For every $\varepsilon\in(0,1)$, we have
\begin{align}
&\varepsilon^{H-\frac{1}{2}}\left(\widehat{X}_1^{\varepsilon,H}-x_0\right)=\varepsilon^{H}
\left[\bar{\rho}\int_0^1\sigma(\varepsilon^{H}\widehat{B}_s)dW_s+\rho
\int_0^1\sigma(\varepsilon^{H}\widehat{B}_s)dB_s\right] \nonumber \\
&\stackrel{\tiny law}{=}\varepsilon^{H}\left[\bar{\rho}
\left\{\int_0^1\sigma(\varepsilon^{H}\widehat{B}_s)^2ds\right\}^{\frac{1}{2}}W_1
+\rho\int_0^1\sigma(\varepsilon^{H}\widehat{B}_s)dB_s\right].
\label{E:li}
\end{align}
The stochastic integrals in (\ref{E:li}) exist, while the ordinary integral exists almost surely, by the $L^2$-condition in (\ref{E:squar}). To prove the equality in law in (\ref{E:li}), we show that the distribution functions of the second and the third expressions in (\ref{E:li}) coincide. This can be done by conditioning on the path of the process $s\mapsto\sigma(\varepsilon^{H}\widehat{B}_s)$ and the value of the random variable $\int_0^1\sigma(\varepsilon^{H}\widehat{B}_s)dB_s$, and taking into account 
the independence of the processes $B$ and $W$. 

It is not hard to prove, using the same ideas as in the proof of Lemma B-1 in \cite{FZ}, that the two-dimensional process $t\mapsto(B_t,\widehat{B}_t)$ is Gaussian. In \cite{FZ}, Lemma B-1 concerns the process $t\mapsto(B_t,B^H_t)$, where $B^H$ is fBm with Hurst parameter $H$. However, the self-similarity property of the process $B^H$ is never used in the proof of Lemma B-1 in \cite{FZ}. 

Let us consider a centered Gaussian vector $G$ in the space $E=C[0,1]^2$ given by $G=(B,\widehat{B})$. The covariance operator $K:E^{*}\mapsto E$ associated with $G$ is as follows:
$$
K(\alpha_1,\alpha_2)(s)=\left(k_1(s),k_2(s)\right),
\quad 0\le s\le 1,
$$
where
$$
k_1(s)=\int_0^1(t\wedge s)d\alpha_1(t)+\int_0^1d\alpha_2(t)\int_0^{t\wedge s}K(t,u)du
$$
and
$$
k_2(s)=\int_0^1d\alpha_1(t)\int_0^{t\wedge s}K(s,u)du+\int_0^1d\alpha_2(t)\int_0^{t\wedge s}K(t,u)K(s,u)du.
$$
Here $(\alpha_1,\alpha_2)\in E^{*}$ is a pair of signed Borel measures of bounded variation on the interval $[0,1]$ .

Our next goal is to apply a well-known large deviation result for Gaussian measures (see Theorem 3.4.12 in \cite{DS}, or Theorem 2.2.3 in \cite{BBK}) to the vector $G$ defined above. We will first define the abstract Wiener space
$\mathcal{W}$ corresponding to $G$ (see Definition 2.2.2 in \cite{BBK}). We set $\mathcal{W}=(W,H,j,\mu)$, where
$H=L^2[0,1]$; 
$
j(f)(t)=\left(\int_0^tf(u)du,\int_0^tK(t,u)f(u)du\right)
$
for $0\le t\le 1$; $W=\overline{j(H)}$, where the closure is taken in the space $C_0[0,1]^2$; 
and $\mu$ is the Gaussian measure on $(W,\mathcal{B}(W))$ induced by the process $t\mapsto\left(B_t,\widehat{B}_t\right)$. It is not hard to see that the space $W$ is a separable Banach space, the space $H$ is a separable Hilbert space, and $j:H\mapsto W$ is a continuous linear injection
(the first component of $j$ is injective). It remains to prove that 
\begin{equation}
\int_W\exp\{i\langle w^{*},w\rangle\}\mu(dw)=\exp\{-\frac{1}{2}||j^{*}w^{*}||_H^2\},
\label{E:Geu}
\end{equation}
where $\langle,\rangle$ denotes the duality between $W^{*}$ and $W$ and $j^{*}:W^{*}\mapsto H^{*}=H$ is the adjoint transformation of $j$.

The space $W^{*}$ is the quotient space of the space of pairs of signed Borel measures of bounded variation on $[0,1]$ by the subspace $A$ of $E^{*}$ annihilating $W$. The annihilation condition is as follows:
$$
(\beta_1,\beta_2)\in A\Longleftrightarrow \int_u^1d\beta_1(t)+\int_u^1K(t,u)d\beta_2(t)=0,
$$
and it is not hard to prove that 
$$
j^{*}(\alpha_1+\beta_1,\alpha_2+\beta_2)(t)=\int_t^1d\alpha_1(s)+\int_t^1K(s,t)d\alpha_2(s),\quad 0\le t\le 1,
$$
where $(\alpha_1,\alpha_2)\in E^{*}$ and $(\beta_1,\beta_2)\in A$. Next, using the known expression for the characteristic function of a Gaussian measure, we get
\begin{equation}
\int_W\exp\{i\langle w^{*},w\rangle\}\mu(dw)=\exp\{-\frac{1}{2}w^{*}(Cw^{*})\}
\label{E:geus}
\end{equation}
where $C:W^{*}\mapsto W$ is the covariance operator. Now, it is clear that in order to prove the 
equality in (\ref{E:Geu}), it suffices to show that 
\begin{equation}
||j^{*}w^{*}||_H^2=w^{*}(Cw^{*}).
\label{E:show}
\end{equation}

Let $w^{*}=(\alpha_1+\beta_1,\alpha_2+\beta_2)$. Then
\begin{align*}
||j^{*}w^{*}||_H^2&=\int_0^1\left(\int_t^1d(\alpha_1+\beta_1)(s)+\int_t^1K(u,t)d(\alpha_2+\beta_2)(u)\right)^2dt
\nonumber \\
&=\int_0^1dt\int_t^1d(\alpha_1+\beta_1)(s)\int_t^1d(\alpha_1+\beta_1)(r) \\
&\quad+2\int_0^1dt\int_t^1d(\alpha_1+\beta_1)(s)
\int_t^1K(u,t)d(\alpha_2+\beta_2)(u) \\
&\quad+\int_0^1dt\int_t^1K(u,t)d(\alpha_2+\beta_2)(u)\int_t^1K(v,t)d(\alpha_2+\beta_2)(v) \\
&=\int_0^1\int_0^1(s\wedge r)d(\alpha_1+\beta_1)(s)d(\alpha_1+\beta_1)(r) \\
&\quad+2\int_0^1\int_0^1d(\alpha_1+\beta_1)(s)d(\alpha_2+\beta_2)(u)\int_0^{s\wedge u}K(u,t)dt
\\
&\quad+\int_0^1\int_0^1d(\alpha_2+\beta_2)(u)d(\alpha_2+\beta_2)(v)\int_0^{u\wedge v}K(u,t)K(v,t)dt.
\end{align*}
Moreover,
\begin{align*}
w^{*}(Cw^{*})&=\mathbb{E}\left[\left(\int_0^1B_sd(\alpha_1+\beta_1)(s)
+\int_0^1\widehat{B}_ud(\alpha_2+\beta_2)(u)\right)^2\right]  \\
&=\int_0^1\int_0^1d(\alpha_1+\beta_1)(s)d(\alpha_1+\beta_1)(r)\mathbb{E}\left[B_sB_r\right]  \\
&\quad+2\int_0^1\int_0^1d(\alpha_1+\beta_1)(s)d(\alpha_2+\beta_2)(u)\mathbb{E}\left[B_s\widehat{B}_u\right] 
\\
&\quad+\int_0^1\int_0^1d(\alpha_2+\beta_2)(u)d(\alpha_2+\beta_2)(v)\mathbb{E}\left[\widehat{B}_u\widehat{B}_v\right]. 
\end{align*}
Next, using the equalities $\mathbb{E}\left[B_sB_r\right]=s\wedge r$; $\mathbb{E}\left[B_s\widehat{B}_u\right]
=\int_0^{s\wedge u}K(u,t)dt$;
and $\mathbb{E}\left[\widehat{B}_u\widehat{B}_v\right]=\int_0^{u\wedge v}K(u,t)K(v,t)dt$, we obtain (\ref{E:show}).
Therefore, $\mathcal{W}$ is an abstract Wiener space.

It follows from Theorem 3.4.12 and Lemma 3.4.2 in \cite{DS} (see also Theorem 2.3 in \cite{BBK}) that the process 
$\varepsilon\mapsto \left(\varepsilon^{H}B,\varepsilon^{H}\widehat{B}\right)$ with state space $C_0[0,1]^2$ satisfies the large deviation principle
with speed $\varepsilon^{-2H}$ and good rate function $I$ defined on the space $C_0[0,1]^2$ as follows.
Recall that for $f\in H_0^1[0,1]$, we defined the function $\widehat{f}$ by (\ref{E:kernel}).
Consider the pairs 
$
(f,\widehat{f})\in C_0[0,1]^2
$ 
for all $f\in H_0^1[0,1]$,
and denote the space of such pairs by ${\cal H}^2$. If $(f,g)\in{\cal H}^2$, we put $I(f,g)=\frac{1}{2}\int_0^1\dot{f}(s)^2ds$, 
while if $(f,g)\in C_0[0,1]^2\backslash{\cal H}^2$, 
we put $I(f,g)=\infty$. 
\begin{remark}\label{R:grf1}
The proof of the above-mentioned fact that $I$ is a good rate function can be found in Section 3.4 of \cite{DS} (see Lemma 3.4.2 in \cite{DS}).
\end{remark}

Since the processes $W$ and $B$ are independent, the process $\varepsilon\mapsto\left(\varepsilon^H W_1,\varepsilon^{H}B,\varepsilon^{H}\widehat{B}\right)$ 
with state space $\mathbb{R}\times C_0[0,1]^2$ satisfies the large deviation principle with speed $\varepsilon^{-2H}$
and rate function $\widetilde{I}(y,f,g)=\frac{1}{2}y^2+I(f,g)$.

Our next goal is to use the previous statement and the extended contraction principle (Theorem 4.2.23 in \cite{DZ}).
Let $\Phi$ be a functional on the space 
$
M=\mathbb{R}\times C_0[0,1]^2
$ 
given by
\begin{equation}
\Phi(y,f,\widehat{f})=\bar{\rho}
\left\{\int_0^1\sigma(\widehat{f}(s))^2ds\right\}^{\frac{1}{2}}y+\rho
\int_0^1\sigma(\widehat{f}(s))\dot{f}(s)ds,
\label{E:lab}
\end{equation}
if $(f,\widehat{f})\in{\cal H}^2$, and by $\Phi(y,f,g)=0$, if $(f,g)\in C_0[0,1]^2\backslash{\cal H}^2$ (compare with the first formula in the proof of Lemma B.3 in \cite{FZ}). Since for $f\in H_0^1[0,1]$, we have $\widehat{f}\in C_0[0,1]$, the functional $\Phi$ is finite on $M$.

For every integer $m\ge 1$, define a functional on $M$ by
\begin{align}
\Phi_m(y,h,l)&=\bar{\rho}\left\{\int_0^1\sigma(l(s))^2ds\right\}^{\frac{1}{2}}y \nonumber \\
&\quad+\rho\sum_{k=0}^{m-1}
\sigma\left(l\left(\frac{k}{m}\right)\right)\left[h\left(\frac{k+1}{m}\right)-h\left(\frac{k}{m}\right)\right].
\label{E:app}
\end{align}
It is clear that for every $m\ge 1$, $\Phi_m:M\mapsto\mathbb{R}$ is a continuous functional.

We will first establish the validity of the condition in formula (4.2.24) in \cite{DZ}.
\begin{lemma}\label{L:ac}
For every $\alpha> 0$,
\begin{align}
&\limsup_{m\rightarrow\infty}\sup_{\left\{f\in H_0^1:\frac{1}{2}y^2
+\frac{1}{2}\int_0^1\dot{f}(s)^2ds\le\alpha\right\}}
|\Phi(y,f,\widehat{f})-\Phi_m(y,f,\widehat{f})|
=0.
\label{E:ecp}
\end{align}
\end{lemma}

\it Proof. \rm Denote $D_{\beta}=\{f\in H_0^1:\int_0^1\dot{f}(s)^2ds<\beta\}$. It is not hard to see that to prove (\ref{E:ecp}), 
it suffices to show that for all $\beta> 0$,
\begin{equation}
\limsup_{m\rightarrow\infty}\left[\sup_{f\in D_{\beta}}\left|J_m(f)
\right|\right]
=0,
\label{E:tru}
\end{equation}
where
\begin{align*}
J_m(f)&=\int_0^1\sigma(\widehat{f}(s))\dot{f}(s)ds 
-\sum_{k=0}^{m-1}
\sigma\left(\widehat{f}\left(\frac{k}{m}\right)\right)\left[f\left(\frac{k+1}{m}\right)
-f\left(\frac{k}{m}\right)\right].
\end{align*}

Set 
$$
h_m(s)=\sum_{k=0}^{m-1}\sigma\left(\widehat{f}\left(\frac{k}{m}\right)\right)
\chi_{\{\frac{k}{m}\le s\le\frac{k+1}{m}\}}.
$$
Then
$
J_m(f)=\int_0^1[\sigma(\widehat{f}(s))-h_m(s)]\dot{f}(s)ds,
$
and hence
\begin{equation}
\sup_{f\in D_{\beta}}\left|J_m(f)
\right|\le\sqrt{\beta}\sup_{f\in D_{\beta}}\sup_{t\in[0,1]}|\sigma(\widehat{f}(t))-h_m(t)|.
\label{E:s1}
\end{equation}
 Since Lemma \ref{L:hra} holds, the set $E_{\beta}=
\{\widehat{f}: f\in D_{\beta}\}$
is precompact in $C_0[0,1]$. It follows from the Arzel\`{a}-Ascoli theorem that the set $E_{\beta}$ 
is uniformly bounded
and equicontinuous in $C_0[0,1]$. Therefore
\begin{equation}
r_{\beta}=\sup_{f\in D_{\beta}}||\widehat{f}||_{C_0[0,1]}<\infty
\label{E:uu1}
\end{equation}
and
\begin{equation}
\omega_{\beta}(\delta)=\sup_{f\in D_{\beta}}\sup_{t_1,t_2\in[0,1]:|t_1-t_2|\le\delta}
|\widehat{f}(t_1)-\widehat{f}(t_2)|\rightarrow 0
\label{E:uu2}
\end{equation}
as $\delta\downarrow 0$. Next, using the local $\omega$-continuity condition for $\sigma$, the estimate in (\ref{E:s1}), 
and the definitions in (\ref{E:uu1}) and (\ref{E:uu2}), we obtain
\begin{equation}
\sup_{f\in D_{\beta}}\left|J_m(f)
\right|\le\sqrt{\beta}L\left(r_{\beta}\right)\omega\left(\omega_{\beta}\left(\frac{1}{m}\right)\right).
\label{E:lef}
\end{equation}
Now, it is clear from (\ref{E:uu1}), (\ref{E:uu2}), and (\ref{E:lef}) that (\ref{E:tru}) holds. This establishes the approximation condition in formula (4.2.24) in \cite{DZ}.

The proof of Lemma \ref{L:ac} is thus completed.

We will next prove that the process $\varepsilon\mapsto \Phi_m\left(\varepsilon^{H}W_1,\varepsilon^{H}B,\varepsilon^{H}\widehat{B}\right)$ 
is an exponentially good approximation as $m\rightarrow\infty$ to the process 
\begin{equation}
\varepsilon\mapsto V_{\varepsilon}=\varepsilon^{H}\left[\bar{\rho}
\left\{\int_0^1\sigma(\varepsilon^{H}\widehat{B}_s)^2ds\right\}^{\frac{1}{2}}W_1
+\rho\int_0^1\sigma(\varepsilon^{H}\widehat{B}_s)dB_s\right] 
\label{E:V}
\end{equation}
(see Definition 4.2.14 in \cite{DZ} for more details on exponentially good approximations).
\begin{lemma}\label{L:ega}
For every $\delta> 0$,
\begin{align}
&\lim_{m\rightarrow\infty}\limsup_{\varepsilon\downarrow 0}\varepsilon^{2H} 
\log\mathbb{P}\left(\left|V_{\varepsilon}-
\Phi_m\left(\varepsilon^{H}W_1,\varepsilon^{H}B,\varepsilon^{H}\widehat{B}\right)\right|>\delta\right)
=-\infty.
\label{E:pust}
\end{align}
\end{lemma}

\it Proof. \rm
Using (\ref{E:app}), we see that in order to prove (\ref{E:pust}), it suffices to show that
\begin{align}
&\lim_{m\rightarrow\infty}\limsup_{\varepsilon\downarrow 0}\varepsilon^{2H}
\log\mathbb{P}\left(\varepsilon^{H}|\rho|\left|\int_0^1\sigma_t^{(m)}dB_t\right|>\delta\right)
=-\infty,
\label{E:liui}
\end{align}
where 
$
\sigma_t^{(m)}=\sigma\left(\varepsilon^{H}\widehat{B}_t\right)
-\sigma\left(\varepsilon^{H}\widehat{B}_{\frac{\left[mt\right]}{m}}\right),
$
for all $0\le t\le 1$. In the previous equality, the symbol $[a]$ stands for the integer part of the number $a\in\mathbb{R}$.

In the sequel, we borrow some ideas from the proof in Appendix B.2 of \cite{FZ}. However, we adapt that proof to our environment, and also need to change certain parts of the proof in \cite{FZ}, since stronger restrictions on the function $\sigma$ and the process $\widehat{B}$ are imposed in \cite{FZ}, than in the present paper. We will establish 
the following stronger condition than that in (\ref{E:liui}):
\begin{align}
&\lim_{m\rightarrow\infty}\limsup_{\varepsilon\downarrow 0}\varepsilon^{2H}
\log\mathbb{P}\left(\varepsilon^{H}|\rho|\sup_{t\in[0,1]}\left|\int_0^t\sigma_s^{(m)}dB_s\right|>\delta\right)
=-\infty.
\label{E:liuis}
\end{align}

It is not hard to see that there exists a positive function $q(\eta)$, where $\eta\in(0,\eta_0)$, for which 
\begin{equation}
q(\eta)\uparrow\infty,\quad L(q(\eta))\uparrow\infty,\quad\mbox{and}\quad L(q(\eta))\omega(\eta)\downarrow 0
\label{E:maps}
\end{equation}
as $\eta\downarrow 0$. Here $L$ is the function appearing in Definition \ref{D:moc}. For instance, we can take 
$
q(\eta)=L^{-1}\left(\widetilde{\omega}(\eta)\right)$ and $\eta_0=\widetilde{\omega}^{-1}(L(0))$,
where $\widetilde{\omega}$ is a strictly decreasing positive continuous function on $(0,\infty)$ such that $\widetilde{\omega}(\eta)\uparrow\infty$ and $\widetilde{\omega}(\eta)\omega(\eta)\downarrow 0$ as $\eta\downarrow 0$.

Fix an integer $m\ge 1$ and a number $\eta$ with $0<\eta<\eta_0$, and define a random variable by 
$$
\xi_{\eta}^{(m)}=\inf\left\{t\in[0,1]:\varepsilon^{H}\left[\frac{\eta}{q(\eta)}
|\widehat{B}_t|+\left|\widehat{B}_t-\widehat{B}_{{\frac{\left[mt\right]}{m}}}\right|\right]>\eta\right\}.
$$
If for $\omega\in\Omega$, the set in the previous definition is empty, then we set $\xi_{\eta}^{(m)}(\omega)=1$. 
It is not hard to see that $\xi_{\eta}^{(m)}$ is a $\{\widetilde{\mathcal{F}}_t\}$-stopping time. Here we use the fact that
the filtration $\{\widetilde{\mathcal{F}}_t\}$ is right-continuous.

Suppose $0< t<\xi_{\eta}^{(m)}$. Then $\frac{\left[mt\right]}{m}<\xi_{\eta}^{(m)}$, and hence
\begin{equation}
\varepsilon^{H}\max\left\{|\widehat{B}_t|,\left|\widehat{B}_{\frac{\left[mt\right]}{m}}\right|\right\}
\le q(\eta)
\label{E:d1}
\end{equation}
and
\begin{equation}
\varepsilon^{H}\left|\widehat{B}_t-\widehat{B}_{{\frac{\left[mt\right]}{m}}}\right|\le\eta.
\label{E:d2}
\end{equation}
Next, using (\ref{E:d1}), (\ref{E:d2}), and the local $\omega$-continuity condition for $\sigma$, we see that
\begin{equation}
|\sigma_t^{(m)}|\le L(q(\eta))\omega(\eta)\quad\mbox{for all}\quad t\in\left[0,\xi_{\eta}^{(m)}\right].
\label{E:sk}
\end{equation}

For every $m\ge 1$, the process $M(t)=\int_0^t\sigma_s^{(m)}dZ_s$ is a local martingale on $[0,1]$. 
We will prove next that the stopped process
\begin{equation}
t\mapsto\int_0^{t\wedge\xi_{\eta}^{(m)}}\sigma_s^{(m)}dZ_s
\label{E:ohh}
\end{equation}
is a martingale. The proof is rather standard, but we decided to include it for the sake of completeness. 

Let $\tau_n\uparrow T$, $n\rightarrow\infty$, be a localizing sequence of stopping times for the process $M$. Then
for every $n\ge 1$, the process $t\mapsto M(t\wedge\tau_n)$ is a martingale, and hence the process 
$M_n(t)=M(t\wedge\tau_n\wedge\xi_{\eta}^{(m)})$ is also a martingale (see Corollary 3.6 in \cite{RY}). Therefore 
for all $0\le s\le t\le T$,
\begin{equation}
\mathbb{E}\left[M_n(t)|{\cal F}_s\right]=M_n(s).
\label{E:l1}
\end{equation}
By the continuity of the sample paths of the process $M$, the expression on the right-hand side of (\ref{E:l1}) 
tends to $M(s\wedge\xi_{\eta}^{(m)})$
as $n\rightarrow\infty$. Our next goal is to pass to the limit as $n\rightarrow\infty$ under the expectation sign on the 
left-hand side of the equality in (\ref{E:l1}). To do that, it suffices to prove the inequality 
\begin{equation}
\mathbb{E}\left[\sup_{n\ge 1}|M_n(t)|\right]<\infty,
\label{E:l2}
\end{equation}
and then use the dominated convergence theorem. Denote by $[M_n]$ the quadratic variation of the process $M_n$. 
Using Doob's maximal inequality,
the definition of quadratic variation, and (\ref{E:sk}), we obtain
\begin{align*}
&\mathbb{E}\left[\sup_{0\le u\le t}M_n(u)^2\right]\le 4\mathbb{E}\left[M_n(t)^2\right]=4\mathbb{E}\left[[M_n](t)\right]
\\
&\le 4\mathbb{E}\left[\int_0^{\xi_{\eta}^{(m)}}\left(\sigma_s^{(m)}\right)^2ds\right]\le 4L(q(\eta)^2\omega(\eta)^2.
\end{align*}
It follows from the previous estimate and the monotone convergence theorem that
$$
\mathbb{E}\left[\sup_{0\le u\le t}M(u\wedge\xi_{\eta}^{(m)})^2\right]<\infty.
$$
Therefore
$$
\mathbb{E}\left[\sup_{n\ge 1}|M_n(t)|\right]\le\mathbb{E}\left[\sup_{0\le u\le t}|M(u\wedge\xi_{\eta}^{(m)})|\right]
<\infty.
$$
This establishes (\ref{E:l2}) and completes the proof of the fact that the process in (\ref{E:ohh}) is a martingale.

Let us fix $\lambda> 0$. Then, for $0<\varepsilon<\varepsilon_1$, the stochastic exponential
$$
{\cal E}_t=\exp\left\{\lambda\varepsilon^H\int_0^{t\wedge\xi_{\eta}^{(m)}}\sigma_s^{(m)}dZ_s-\frac{1}{2}\lambda^2\varepsilon^{2H}
\int_0^{t\wedge\xi_{\eta}^{(m)}}
\left(\sigma_s^{(m) }\right)^2ds\right\}
$$
is a martingale (use (\ref{E:sk}) and Novikov's condition). We will assume in the rest of the proof that $0<\varepsilon<\varepsilon_1$. 
It follows from (\ref{E:sk}) and the martingality condition formulated above
that
\begin{align}
&\mathbb{E}\left[\exp\left\{\lambda\varepsilon^H\int_0^{t\wedge\xi_{\eta}^{(m)}}\sigma_s^{(m)}dZ_s\right\}\right]
=\mathbb{E}\left[{\cal E}_t\exp\left\{\frac{1}{2}\lambda^2\varepsilon^{2H}
\int_0^{t\wedge\xi_{\eta}^{(m)}}
\left(\sigma_s^{(m)}\right)^2ds\right\}\right]\nonumber \\
&\le\exp\left\{\frac{1}{2}\lambda^2\varepsilon^{2H}L(q(\eta))^2\omega(\eta)^2\right\}<\infty,
\label{E:nak1}
\end{align}
for all $t\in[0,1]$. Plugging $t=1$ into (\ref{E:nak1}), we get
\begin{equation}
\mathbb{E}\left[\exp\left\{\lambda\varepsilon^H\int_0^{\xi_{\eta}^{(m)}}\sigma_s^{(m)}dZ_s\right\}\right]
\le\exp\left\{\frac{1}{2}\lambda^2\varepsilon^{2H}L(q(\eta))^2\omega(\eta)^2\right\}.
\label{E:nak2}
\end{equation}

Since the process in (\ref{E:ohh}) is a martingale, the integrability condition in (\ref{E:nak1}) implies that the process 
$$
t\mapsto\exp\left\{\lambda\varepsilon^H\int_0^{t\wedge\xi_{\eta}^{(m)}}\sigma_s^{(m)}dZ_s\right\}
$$
is a positive submartingale (see Proposition 3.6 in \cite{KaS}). Next, using the first submartingale inequality in 
\cite{KaS}, Theorem 3.8, we obtain
\begin{align*}
&\mathbb{P}\left(\sup_{t\in[0,\xi_{\eta}^{(m)}]}
\exp\left\{\varepsilon^{H}\lambda\int_0^t\sigma_s^{(m)}dB_s\right\}> e^{\lambda\delta}\right) \\
&\le\exp\left\{\frac{1}{2}\varepsilon^{2H}\lambda^2
L(q(\eta))^2\omega(\eta)^{2}-\lambda\delta\right\}.
\end{align*}
Setting $\lambda=\frac{\delta}{\varepsilon^{2H}L(q(\eta))^2\omega(\eta)^{2}}$, we get from the previous inequality that
\begin{align}
&\mathbb{P}\left(\sup_{t\in[0,\xi_{\eta}^{(m)}]}
\varepsilon^{H}\int_0^t\sigma_s^{(m)}dB_s>\delta\right) 
\le\exp\left\{-\frac{\delta^2}{2\varepsilon^{2H}L(q(\eta))^2\omega(\eta)^{2}}\right\}.
\label{E:five}
\end{align}
Next, using the fact that the process $t\mapsto-B_t$ is a Brownian motion, we derive from (\ref{E:five}) that
$$
\mathbb{P}\left(\sup_{t\in[0,\xi_{\eta}^{(m)}]}
\varepsilon^{H}\left|\int_0^t\sigma_s^{(m)}dB_s\right|>\delta\right)
\le 2\exp\left\{-\frac{\delta^2}{2\varepsilon^{2H}L(q(\eta))^2\omega(\eta)^{2}}\right\}.
$$
Now, replacing $\delta$ by $\frac{\delta}{|\rho|}$, transforming the resulting inequality, and using the last statement in (\ref{E:maps}), we obtain the following equality:
\begin{align}
&\lim_{\eta\downarrow 0}\,\sup_{m\ge 1}\,\limsup_{\varepsilon\downarrow 0}\varepsilon^{2H}
\log\mathbb{P}\left(\varepsilon^{H}|\rho|\sup_{t\in[0,\xi_{\eta}^{(m)}]}\left|\int_0^t\sigma_s^{(m)}dB_s\right|>\delta\right)
=-\infty.
\label{E:liuiss}
\end{align}

It is not hard to see that
\begin{align}
&\mathbb{P}\left(\varepsilon^{H}|\rho|\sup_{t\in[0,1]}\left|\int_0^t\sigma_s^{(m)}dB_s\right|>\delta\right)
\le\mathbb{P}\left(\xi_{\eta}^{(m)}< 1\right) \nonumber \\ &+\mathbb{P}\left(\varepsilon^{H}|\rho|\sup_{t\in[0,\xi_{\eta}^{(m)}]}\left|\int_0^t\sigma_s^{(m)}dB_s\right|>\delta\right)
\label{E:pom}
\end{align}
and
\begin{align}
\mathbb{P}\left(\xi_{\eta}^{(m)}< 1\right)&\le\mathbb{P}\left(\sup_{t\in[0,1]}\varepsilon^{H}
\left[\frac{\eta}{q(\eta)}
|\widehat{B}_t|+\left|\widehat{B}_t-\widehat{B}_{{\frac{\left[mt\right]}{m}}}\right|\right]>\eta\right) \nonumber \\
&\le\mathbb{P}\left(\sup_{t\in[0,1]}
|\widehat{B}_t|>\frac{q(\eta)}{2\varepsilon^{H}}\right)
+\mathbb{P}\left(\sup_{t\in[0,1]}\varepsilon^{H}\left|\widehat{B}_t-\widehat{B}_{{\frac{\left[mt\right]}{m}}}\right|
>\frac{\eta}{2}\right).
\label{E:dob}
\end{align}

We will need the following auxiliary statement.
\begin{lemma}\label{L:byl}
For every $y> 0$,
\begin{equation}
\limsup_{m\rightarrow\infty}\limsup_{\varepsilon\downarrow 0}\varepsilon^{2H}\log\mathbb{P}\left(\sup_{t\in[0,1]}\varepsilon^{H}\left|\widehat{B}_t-\widehat{B}_{{\frac{\left[mt\right]}{m}}}\right|>y\right)
=-\infty.
\label{E:ma1}
\end{equation}
\end{lemma}
\begin{remark}\label{R:liem}
For fractional Brownian motion, Lemma \ref{L:byl} was established in \cite{FZ} (see the proof on p. 138 of \cite{FZ}). The authors of \cite{FZ} used the fact that fBm has stationary increments. Gaussian processes in Theorem \ref{T:1} do not necessarily have stationary increments. However, Lemma \ref{L:byl} still holds. 
\end{remark}

\it Proof. \rm We have
\begin{align}
&\mathbb{P}\left(\sup_{t\in[0,1]}\left|\widehat{B}_t-\widehat{B}_{{\frac{\left[mt\right]}{m}}}\right|
>\varepsilon^{-H}y\right)
\le\mathbb{P}\left(\sup_{t_1,t_2\in[0,1]:|t_1-t_2|\le\frac{1}{m}}\left|\widehat{B}_{t_1}-\widehat{B}_{t_2}\right|>\varepsilon^{-H}y\right).
\label{E:cur}
\end{align}

To estimate the term on the right-hand side of (\ref{E:cur}), we will use estimates obtained in an interesting paper \cite{CC} of Cs\'{a}ki and Cs\"{o}rg\H{o}.
The estimates in \cite{CC} are formulated for continuous stochastic processes indexed by the real line. With no loss of generality, we may assume that $\widehat{B}_t=0$ for $t< 0$ and $\widehat{B}_t=\widehat{B}_1$ if $t> 1$. Let $h> 0$ be a small number and let $t\in[0,1-h]$. Recall that  $\widehat{B}$ is a Volterra type process. It follows from (\ref{E:ooo}) that 
$\widehat{B}_{t+h}-\widehat{B}_t$ is a Gaussian random variable with mean zero and variance $V(t,h)=\int_0^1|K(t+h,s)-K(t,s)|^2ds$. Therefore, we have $V(t,h)\le c\,h^{\alpha}$, $h< h_0$, $t\in[0,1-h]$. 
It follows that there exists $x^{*}> 0$ such that for all $x> x^{*}$,
\begin{align}
&\mathbb{P}\left(\left|\widehat{B}_{t+h}-\widehat{B}_t\right|>xh^{\frac{\alpha}{2}}\right)
=\frac{1}{\sqrt{2\pi V(t,h)}}\int_{xh^{\frac{\alpha}{2}}}^{\infty}\exp\left\{-\frac{y^2}{2V(t,h)}\right\}dy 
\nonumber \\
&=\frac{1}{\sqrt{2\pi}}\int_{\frac{xh^{\frac{\alpha}{2}}}{\sqrt{V(t,h)}}}^{\infty}\exp\left\{-\frac{z^2}{2}\right\}dz
\le\frac{1}{\sqrt{2\pi}}\int_{\frac{x}{\sqrt{c}}}^{\infty}\exp\left\{-\frac{z^2}{2}\right\}dz \nonumber \\
&\le\frac{\sqrt{2c}}{\sqrt{\pi}x}\exp\left\{-\frac{x^2}{2c}\right\}
\le\frac{\sqrt{2c}}{\sqrt{\pi}x^{*}}\exp\left\{-\frac{x^2}{2c}\right\}.
\label{E:cc}
\end{align}
Next, using (\ref{E:cc}), we see that the estimate in (2.1) in \cite{CC} holds with $K=\frac{\sqrt{2c}}{\sqrt{\pi}x^{*}}$, $\gamma=\frac{1}{2c}$, $\beta=2$, and
$\sigma(h)=h^{\frac{\alpha}{2}}$.

Now, we are ready to use Lemma 2.2 in \cite{CC}.
It follows from this lemma with $L(h)=1$ and $\frac{\alpha}{2}$ instead of $\alpha$ that there exist constants $C> 0$, $\delta> 0$, $h_0> 0$, and $x_0> 0$ such that
\begin{align}
&\mathbb{P}\left(\sup_{t\in[0,1-h]}\sup_{s\in(0,h)}
\left|\widehat{B}_{t+s}-\widehat{B}_t\right|>xh^{\frac{\alpha}{2}}\right) 
\le\frac{C}{h}\exp\left\{-\delta x^2\right\},
\label{E:neja}
\end{align}
for all $x> x_0$ and $0< h< h_0$.

Our next goal is to derive (\ref{E:ma1}) from (\ref{E:neja}). Let us take any integer $m$ such that $\frac{1}{m}< h_0$,
and put $h=\frac{1}{m}$. Let us also take any $\varepsilon> 0$ such that the number $x$ defined by $x=m^{\frac{\alpha}{2}}\varepsilon^{-H}y$, where $y$ is a number in Lemma \ref{L:byl}, satisfies $x> x_0$. Then, applying (\ref{E:neja}), we obtain
\begin{align}
&\mathbb{P}\left(\sup_{t_1,t_2\in[0,1]:|t_1-t_2|\le\frac{1}{m}}\left|\widehat{B}_{t_1}-\widehat{B}_{t_2}\right|>\varepsilon^{-H}y\right)
\le Cm\exp\left\{-\delta m^{\alpha}\varepsilon^{-2H}y^2\right\}.
\label{E:pri}
\end{align}
It is easy to see that (\ref{E:ma1}) follows from (\ref{E:pri}).

This completes the proof of Lemma \ref{L:byl}.

Let us return to the proof of Lemma \ref{L:ega}.
It is not hard to see using (\ref{E:addik}) that
$$
\limsup_{\eta\downarrow 0}\limsup_{\varepsilon\downarrow 0}\varepsilon^{2H}\log\mathbb{P}\left(\sup_{t\in[0,1]}
|\widehat{B}_t|>\frac{q(\eta)}{2\varepsilon^{H}}\right)=-\infty.
$$
It follows from (\ref{E:liuiss}) that for every large number $N> 0$ there exists a small number $\eta> 0$, depending on $N$ and such that
\begin{align}
&\sup_{m\ge 1}\,\limsup_{\varepsilon\downarrow 0}\varepsilon^{2H}
\log\mathbb{P}\left(\varepsilon^{H}|\rho|\sup_{t\in[0,\xi_{\eta}^{(m)}]}\left|\int_0^t\sigma_s^{(m)}dB_s\right|>\delta\right)<-N
\label{E:stro}
\end{align}
and 
\begin{align}
&\limsup_{\varepsilon\downarrow 0}\varepsilon^{2H}\log\mathbb{P}\left(\sup_{t\in[0,1]}
|\widehat{B}_t|>\frac{q(\eta)}{2\varepsilon^{H}}\right)<-N.
\label{E:stroj}
\end{align}

Let $N$ and $\eta$ be numbers satisfying (\ref{E:stro}) and (\ref{E:stroj}). Then, using the inequality
$$
\log(a+b)\le\max(\log(2a),\log(2b)),\quad a> 0,\,\,b> 0,
$$ 
and taking into account (\ref{E:dob}), (\ref{E:ma1}), and (\ref{E:stroj}),
we see that there exists $m_0> 1$, depending on $N$, and such that the following condition holds: For every $m\ge m_0$ there exists
$\varepsilon_m> 0$, depending on $m$ and $\eta$, and satisfying
\begin{align}
\varepsilon^{2H}\log\mathbb{P}\left(\xi_{\eta}^{(m)}< 1\right)\le
-N,
\label{E:alik}
\end{align}
for all $\varepsilon\in(0,\varepsilon_m]$. Next, using 
(\ref{E:pom}), (\ref{E:stro}) and (\ref{E:alik}), and reasoning as above, we see that
\begin{align}
&\limsup_{m\rightarrow\infty}\,\limsup_{\varepsilon\downarrow 0}\varepsilon^{2H}\log\mathbb{P}\left(\varepsilon^{H}|\rho|\sup_{t\in[0,1]}\left|\int_0^t\sigma_s^{(m)}dB_s\right|>\delta\right)<-N.
\label{E:pomo}
\end{align}
Since $N$ can be any large positive number, (\ref{E:pomo}) implies (\ref{E:liuis}).

This completes the proof of Lemma \ref{L:ega}.

Finally, by taking into account (\ref{E:ecp}), (\ref{E:V}), and (\ref{E:pust}), and applying the extended contraction principle formulated in Theorem 4.2.23 in \cite{DZ}, we see that the process $\varepsilon\mapsto V_{\varepsilon}$ (see (\ref{E:V}) for the definition of $V_{\varepsilon}$) satisfies the large deviation principle with speed  $\varepsilon^{-2H}$ and good rate function $I_1$ given by
\begin{equation}
I_1(x)=\inf_{f\in H_0^1[0,1]}\left[\frac{1}{2}y^2+\frac{1}{2}\int_0^1\dot{f}(s)^2ds:\Phi(y,f,\widehat{f})=x\right],
\label{E:vu}
\end{equation}
where the function $\Phi$ is defined in (\ref{E:lab}). 
Next, recalling that for every $\varepsilon$, 
$$
V_{\varepsilon}=\varepsilon^{H-\frac{1}{2}}\left(\widehat{X}^{\varepsilon,H}_1-x_0\right)
$$
in law (see (\ref{E:li})), we prove that the same LDP holds for the process $\varepsilon\mapsto\varepsilon^{H-\frac{1}{2}}\left(\widehat{X}^{\varepsilon,H}_1-x_0\right)$.
\begin{remark}\label{R:grf2}
The fact that $I_1$ is a good rate function follows from Remark \ref{R:grf1} and Theorem 4.2.23 in \cite{DZ}.
\end{remark}

It follows from (\ref{E:lab}) and (\ref{E:vu}) that for every $f\in H_0^1[0,1]$, $y$ should be equal to 
$$
\frac{x-\rho\int_0^1\sigma(\widehat{f}(s))\dot{f}(s)ds}{\bar{\rho}\left\{\int_0^1\sigma(\widehat{f}(s))^2ds\right\}^{\frac{1}{2}}}.
$$
Now, using (\ref{E:vu}), we obtain formula (\ref{E:vunz}).

This completes the proof of Theorem \ref{T:1} in the case where $T=1$.
 
\section{Asymptotic behavior of option pricing functions and the implied volatility}\label{S:sta}
Let us recall that for every $\varepsilon\in(0,1]$, we denoted by $S_t^{\varepsilon,H}$, $0\le t\le T$, the process 
satisfying (\ref{E:scaled}). The next statement concerns small-noise behavior of 
binary options. The continuity and monotonicity properties of the function $I_T$ established in Lemma \ref{L:conti}
play an important role in the proof of Corollary \ref{C:bob} and the other assertions below.
\begin{corollary}\label{C:bob}
Let $H> 0$, and suppose the conditions in Theorem \ref{T:1} hold. Then, the following equalities hold:
\begin{equation}
\lim_{\varepsilon\downarrow 0}\varepsilon^{2H}\log\mathbb{P}\left(S_T^{\varepsilon,H}
>s_0\exp\left\{y\varepsilon^{\frac{1}{2}-H}\right\}\right)=-I_T(y),
\quad y> 0,
\label{E:vv1}
\end{equation}
and
\begin{equation}
\lim_{\varepsilon\downarrow 0}\varepsilon^{2H}\log\mathbb{P}\left(S_T^{\varepsilon,H}
< s_0\exp\left\{y\varepsilon^{\frac{1}{2}-H}\right\}\right)=-I_T(y),\quad y< 0.
\label{E:vvv1}
\end{equation}
\end{corollary}

\it Proof. \rm Using Theorem \ref{T:1} and Lemma \ref{L:conti}, we see that for all $y> 0$,
\begin{equation}
\lim_{\varepsilon\downarrow 0}\varepsilon^{2H}\log\mathbb{P}\left(\varepsilon^{H-\frac{1}{2}}
\left(X_T^{\varepsilon,H}-x_0\right)> y\right)=-I_T(y).
\label{E:v2}
\end{equation}
Now, it is easy to prove that (\ref{E:vv1}) follows from (\ref{E:v2}). The proof of (\ref{E:vvv1}) is similar.
\begin{remark}\label{R:atm}
In the case where $0< H<\frac{1}{2}$, we have
$
s_0\exp\left\{y\varepsilon^{\frac{1}{2}-H}\right\}\rightarrow s_0
$
as $\varepsilon\downarrow 0$. Hence, for $H<\frac{1}{2}$, formula
(\ref{E:vv1}) describes small-noise behavior of a binary option in near-the-money regime. 
Note that we do not assume in Corollary \ref{C:bob} that the process $\widehat{B}$ is self-similar.
\end{remark}

We will next discuss the small-maturity behavior of digital options. Fix a time horizon $\widetilde{T}$, 
and denote by $T\in (0,\widetilde{T})$ the maturity of the option. Let us recall that the function 
$\widehat{I}_{\widetilde{T}}$ is defined in (\ref{E:dop}). The next assertion can be established exactly 
as Corollary \ref{C:bob}, using Theorem \ref{T:smt} instead of Theorem \ref{T:1}. It generalizes Corollary 4.7 in \cite{FZ}. 
\begin{corollary}\label{C:bo}
Let $0< H< 1$, and suppose the conditions in Theorem \ref{T:smt} hold. Suppose also that the process $\widehat{B}$ is $H$-self-similar. Then, the following equalities are valid:
$$
\lim_{T\downarrow 0}T^{2H}\log\mathbb{P}\left(S_T> s_0\exp\left\{yT^{\frac{1}{2}-H}\right\}\right)=-\widehat{I}_{\widetilde{T}}(y),
\quad y> 0
$$
and
$$
\lim_{T\downarrow 0}T^{2H}\log\mathbb{P}\left(S_T< s_0\exp\left\{yT^{\frac{1}{2}-H}\right\}\right)
=-\widehat{I}_{\widetilde{T}}(y),\quad y< 0.
$$
\end{corollary}
\begin{remark}\label{R:too}
For $0< H<\frac{1}{2}$, Corollary \ref{C:bo} characterizes the small-maturity behavior of a binary option
in near-the-money regime (see Remark \ref{R:atm}).
\end{remark}
\begin{remark}\label{R:fui}
The expressions on the left-hand sides of the formulas in Corollary \ref{C:bo} do not depend on the time-horizon $\widetilde{T}$. 
This is in agreement with the fact that for every $0<\widetilde{T}_1\le\widetilde{T}$ and $y> 0$, we have 
$\widehat{I}_{\widetilde{T}_1}(y)=\widehat{I}_{\widetilde{T}}(y)$ (see Remark \ref{R:impo}).
\end{remark}

We will next turn our attention to the small-maturity and the small-noise behavior of call and put pricing functions. We will also study similar problems for the implied volatility. It will be assumed that the time horizon $\widetilde{T}$ is fixed, and the maturity $T$ of the option satisfies $0< T<\widetilde{T}$. By $K> 0$ will be denoted the strike price. Let us also denote by $k$ the log-monenyness given by 
$k=\log\frac{K}{s_0}$. Then, the call and put pricing functions are defined by
$$
C(T,K)=\mathbb{E}\left[(S_T-K)^{+}\right]\quad\mbox{and}\quad P(T,K)=\mathbb{E}\left[(K-S_T)^{+}\right],
$$
respectively. In the small-noise case, we set
$$
C^{H}(\varepsilon,K)=\mathbb{E}\left[(S_T^{\varepsilon,H}-K)^{+}\right]\quad\mbox{and}
\quad P^{H}(\varepsilon,K)=\mathbb{E}\left[(K-S_T^{\varepsilon,H})^{+}\right]
$$

The implied volatility $(T,K)\mapsto\widehat{\sigma}(T,K)$, associated with the call pricing function $C$, is defined as the value of the volatility parameter $\sigma$ in the Black-Scholes model for which
$C(T,K)=C_{BS}(T,K,\sigma=\widehat{\sigma}(T,K))$. The definition of the implied volatility $(\varepsilon,K)\mapsto\widehat{\sigma}^{H}(\varepsilon,K)$, associated with the call pricing function $C^{H}$, is similar.

The next statement concerns the asymptotic behavior of the call and put pricing functions in mixed small-noise and mixed small-maturity regimes. In the small-maturity case, we assume that the process $\widehat{B}$ is $H$-self-similar, and the strike price is $T$-dependent and given by $K_T=s_0\exp\{yT^{\frac{1}{2}-H}\}$. The log-moneyness in this regime satisfies $k_T=yT^{\frac{1}{2}-H}$. In the small-noise case, we do not asssume the self-similarity of the process $\widehat{B}$. In this case, we set 
$K_T^{\varepsilon,H}=s_0\exp\{y\varepsilon^{\frac{1}{2}-H}\}$ and $k_T^{\varepsilon,H}=y\varepsilon^{\frac{1}{2}-H}$.
\begin{corollary}\label{C:call}
(i)\,Let $H> 0$, and suppose the conditions in Theorem \ref{T:1} hold. Suppose also that the linear growth condition in Lemma \ref{L:environ} is satisfied. Then, for every $y> 0$,
\begin{equation}
\lim_{\varepsilon\downarrow 0}\varepsilon^{2H}\log C^{H}\left(\varepsilon,K_T^{\varepsilon,H}\right)=-I_T(y),
\label{E:callo1}
\end{equation}
while for $y< 0$,
\begin{equation}
\lim_{\varepsilon\downarrow 0}\varepsilon^{2H}\log P^{H}\left(\varepsilon,K_T^{\varepsilon,H}\right)=-I_T(y).
\label{E:callo2}
\end{equation}
(ii)\,Let $0< H< 1$, and suppose the conditions in Theorem \ref{T:smt} hold. Suppose also that the process $\widehat{B}$ is $H$-self-similar, and the linear growth condition in Lemma \ref{L:environ} is satisfied. Then, for every $y> 0$,
\begin{equation}
\lim_{T\downarrow 0}T^{2H}\log C(T,K_T)=-\widehat{I}_{\widetilde{T}}(y),
\label{E:call1}
\end{equation}
while for $y< 0$,
\begin{equation}
\lim_{T\downarrow 0}T^{2H}\log P(T,K_T)=-\widehat{I}_{\widetilde{T}}(y).
\label{E:call2}
\end{equation}
\end{corollary}
\begin{remark}\label{R:also}
For $0< H<\frac{1}{2}$, Corollary \ref{C:call} characterizes the small-time near-the-money behavior of call and put prices 
(see Remarks \ref{R:atm} and \ref{R:too}).
\end{remark}
\begin{remark}\label{R:per}
Formula (\ref{E:callo1}) can be rewritten as follows:
\begin{equation}
\log \frac{1}{C^{H}\left(\varepsilon,K_T^{\varepsilon,H}\right)}=\varepsilon^{-2H}I_T(y)+o_y(\varepsilon^{-2H}),
\label{E:call3}
\end{equation}
as $t\downarrow 0$. Formula (\ref{E:callo2}) can be transformed similarly, using the put pricing function 
$P^{H}\left(\varepsilon,K_T^{\varepsilon,T}\right)$, $0<\varepsilon\le 1$.
\end{remark}

\it Proof of Corollary \ref{C:call}. \rm We will only prove part (i) of Corollary \ref{C:call}. The proof of part (ii) is similar. There are well-known methods allowing one to pass from large deviation estimates for the log-price to similar estimates for the call and put pricing functions. The interested reader can consult the papers \cite{FZ,P} (see the proof on p. 36 in \cite{P}, or the proof of Corollary 4.13 in \cite{FZ}). Part 2 of Corollary \ref{C:call} generalizes Corollary 4.13 in \cite{FZ}. 

For $y> 0$, the lower LDP estimate for the call can be obtained as follows. Fix $\delta> 0$. Then we have
\begin{align*}
C^{H}\left(\varepsilon,K_T^{\varepsilon,H}\right)
&\ge s_0\exp\{y\varepsilon^{\frac{1}{2}-H}\}\left[\exp\{\delta \varepsilon^{\frac{1}{2}-H}\}-1\right] \\
&\quad\times\mathbb{P}\left(S_T^{\varepsilon,H}>s_0\exp\{(y+\delta)\varepsilon^{\frac{1}{2}-H}\}\right).
\end{align*} 
Next, applying the $\liminf_{\varepsilon\downarrow 0}\varepsilon^{2H}\log$-transformation to the both sides of the previous estimate, and taking into account the inequality $e^r-1\ge r$ for all $r> 0$, we obtain
\begin{align*}
\liminf_{\varepsilon\downarrow 0}\varepsilon^{2H}\log C^{H}\left(\varepsilon,K_T^{\varepsilon,H}\right)&\ge\liminf_{\varepsilon\downarrow 0}\varepsilon^{2H}\log \mathbb{P}\left(S_T^{\varepsilon,H}>s_0\exp\{(y+\delta)\varepsilon^{\frac{1}{2}-H}\}\right). 
\end{align*}
It follows from the first formula in Corollary \ref{C:bob} with $y+\delta$ instead of $y$ and from the previous estimate that 
\begin{align*}
&\liminf_{\varepsilon\downarrow 0}\varepsilon^{2H}\log C^{H}\left(\varepsilon,K_T^{\varepsilon,H}\right)\ge
-I_T(y+\delta).
\end{align*}
Finally, since the function $I_T$ is continuous (see Lemma \ref{L:conti}), the lower LDP estimate for the call pricing function holds true.

We will next turn out attention to the upper estimate. Let $p> 1$ and $q> 1$ be such that $\frac{1}{p}+\frac{1}{q}=1$.
Then
$$
C^{H}\left(\varepsilon,K_T^{\varepsilon,H}\right)\le\left\{\mathbb{E}\left[\left|S_T^{\varepsilon,H}\right|^p\right]\right\}^{\frac{1}{p}}\left\{
\mathbb{P}\left(S_T^{\varepsilon,H}>s_0\exp\left\{y\varepsilon^{\frac{1}{2}-H}\right\}\right)\right\}^{\frac{1}{q}}.
$$
Using the first formula in Corollary \ref{C:bob}, we obtain
\begin{equation}
\limsup_{\varepsilon\downarrow 0}\varepsilon^{2H}\log C^{H}\left(\varepsilon,K_T^{\varepsilon,H}\right)
\le\frac{1}{p}\limsup_{\varepsilon\downarrow 0}\varepsilon^{2H}\log 
\mathbb{E}\left[\left|S_T^{\varepsilon,H}\right|^p\right]-\frac{1}{q}I_T(y).
\label{E:pre}
\end{equation}

It remains to estimate the expectation on the right-hand side of formula (\ref{E:pre}). 
It follows from (\ref{E:firs}) that
\begin{align}
&\mathbb{E}\left[\left|S_T^{\varepsilon,H}\right|^p\right] \nonumber \\
&=s_0^p\mathbb{E}\left[\exp\left\{-\frac{p\varepsilon}{2}
\int_0^t\sigma^2(\varepsilon^H\widehat{B}_s)ds+p\sqrt{\varepsilon}\bar{\rho}\int_0^t\sigma(\varepsilon^H\widehat{B}_s)dW_s
+p\sqrt{\varepsilon}\rho\int_0^t\sigma(\varepsilon^H\widehat{B}_s)dB_s\right\}\right]
\nonumber \\
&\le s_0^p\left\{\mathbb{E}\left[\exp\left\{-2p^2\varepsilon\int_0^T\sigma^2(\varepsilon^H
\widehat{B}_s)ds+2p\sqrt{\varepsilon}
\bar{\rho}\int_0^T\sigma(\varepsilon^H\widehat{B}_s)dW_s+2p\sqrt{\varepsilon}
\rho\int_0^T\sigma(\varepsilon^H\widehat{B}_s)dB_s\right\}\right]\right\}^{\frac{1}{2}} \nonumber \\
&\times\left\{\mathbb{E}\left[\exp\left\{\left(2p^2-p\right)\varepsilon
\int_0^t\sigma^2(\widehat{B}_s)ds\right\}\right]\right\}^{\frac{1}{2}}.
\label{E:itfol}
\end{align}
By Lemma \ref{L:environ}, for every $\varepsilon> 0$, the process
$$
t\mapsto\exp\left\{-2p^2\varepsilon\int_0^t\sigma^2(\varepsilon^H
\widehat{B}_s)ds+2p\sqrt{\varepsilon}
\bar{\rho}\int_0^t\sigma(\varepsilon^H\widehat{B}_s)dW_s+2p\sqrt{\varepsilon}
\rho\int_0^t\sigma(\varepsilon^H\widehat{B}_s)dB_s\right\}
$$
is a martingale. It follows from (\ref{E:itfol}) that
$$
\mathbb{E}\left[\left|S_T^{\varepsilon,H}\right|^p\right]\le s_0^p
\left\{\mathbb{E}\left[\exp\left\{\left(2p^2-p\right)\varepsilon
\int_0^T\sigma^2(\varepsilon^H\widehat{B}_s)ds\right\}\right]\right\}^{\frac{1}{2}}.
$$
Moreover, using the linear growth condition in Lemma \ref{L:environ}, we obtain
\begin{align*}
&\mathbb{E}\left[\left|S_T^{\varepsilon,H}\right|^p\right]
\le s_0^p\exp\left\{\left(p^2-\frac{p}{2}\right)c_1\varepsilon T\right\}
\left\{\mathbb{E}\left[\exp\left\{\left(2p^2-p\right)c_2\varepsilon^{2H+1}\int_0^T\widehat{B}_s^2ds\right\}\right]\right\}
^{\frac{1}{2}}.
\end{align*}

In the sequel, we will use a weaker version of the previous estimate. This version is simpler to work with, and is sufficient for our purposes. Since $\varepsilon< 1$, the estimate above gives
\begin{align}
&\mathbb{E}\left[\left|S_T^{\varepsilon,H}\right|^p\right]
\le s_0^p\exp\left\{\left(p^2-\frac{p}{2}\right)c_1T\right\}
\left\{\mathbb{E}\left[\exp\left\{\left(2p^2-p\right)c_2\varepsilon\int_0^T\widehat{B}_s^2ds\right\}\right]\right\}
^{\frac{1}{2}}.
\label{E:mard}
\end{align}

Our next goal is to prove an auxiliary statement (Lemma \ref{L:hah} below) that will allow us to estimate the expectation appearing on the right-hand side of (\ref{E:mard}). Our proof of Lemma \ref{L:hah} is qualitatively different from the proof of similar estimates in \cite{FZ}. It uses the Karhunen-Lo\'{e}ve decomposition of the process $\widehat{B}$ (for more information on the Karhunen-Lo\'{e}ve theorem, see, e.g., \cite{Y}). 
It follows from this theorem that there exist a nonincreasing sequence $\lambda_k$, $k\ge 1$, of positive numbers 
with $\sum_{k=1}^{\infty}\lambda_k<\infty$, and an i.i.d. sequence of standard normal variables $Z_k$, $k\ge 1$, such that 
\begin{equation}
\int_0^T\widehat{B}_u^2du=\sum_{k=1}^{\infty}\lambda_kZ_k^2\quad\mbox{in distribution}
\label{E:lew}
\end{equation}
(see, e.g., formula (5) in \cite{GVZ1}). The symbols $\lambda_k$, $k\ge 1$, stand for positive eigenvalues (counting the multiplicities) of the covariance operator associated with the process $\widehat{B}$, and it is assumed that 
the eigenvalues are rearranged in decreasing order. 
\begin{lemma}\label{L:hah}
Let $a> 0$ be a real number. Then, for all 
\begin{equation}
0<\varepsilon<(4a\lambda_1)^{-1},
\label{E:T}
\end{equation}
the following estimate holds:
\begin{equation}
\mathbb{E}\left[\exp\left\{a\varepsilon\int_0^T\widehat{B}_u^2du\right\}\right]\le \exp\left\{2a\sum_{k=1}^{\infty}\lambda_k\right\}.
\label{E:TT}
\end{equation}
\end{lemma}
 
 \it Proof. \rm Using formula (\ref{E:lew}), we obtain
\begin{align*}
 &D=:\mathbb{E}\left[\exp\left\{a\varepsilon\int_0^T\widehat{B}_u^2du\right\}\right]
 \le\mathbb{E}\left[\exp\left\{a\varepsilon\sum_{k=1}^{\infty}\lambda_kZ_k^2\right\}\right]
 \\
 &=\prod_{k=1}^{\infty}\mathbb{E}\left[\exp\left\{a\varepsilon\lambda_kZ_k^2\right\}\right]
=\prod_{k=1}^{\infty}\left(1-2a\varepsilon\lambda_k\right)^{-\frac{1}{2}}.
\end{align*}
Therefore 
\begin{equation}
\log D\le\frac{1}{2}\sum_{k=1}^{\infty}\log\left(1+\frac{2a\varepsilon\lambda_k}
{1-2a\varepsilon\lambda_k}\right). 
\label{E:pomen}
\end{equation}

Suppose (\ref{E:T}) holds. Then $2a\varepsilon\lambda_k<\frac{1}{2}$ for all $k\ge 1$, and hence,
$\frac{2a\varepsilon\lambda_k}{1-2a\varepsilon\lambda_k}< 1$, $k\ge 1$. Next, using (\ref{E:pomen}), 
the previous estimate, and the inequality $\log(1+h)\le h$, $0< h< 1$,
we get
\begin{align*}
&\log D\le a\varepsilon\sum_{k=1}^{\infty}
\frac{\lambda_k}{1-2a\varepsilon\lambda_k}
\le 2a\sum_{k=1}^{\infty}\lambda_k.
\end{align*}
Now, it is clear that the estimate in Lemma \ref{L:hah} holds.

This completes the proof of Lemma \ref{L:hah}.

Let $a=(2p^2-p)c_2$ in Lemma \ref{L:hah}. It is not hard to see, using (\ref{E:TT}) and (\ref{E:mard}), that for all $p> 1$,
$$
\limsup_{\varepsilon\downarrow 0}\varepsilon^{2H}\log 
\mathbb{E}\left[\left|S_T^{\varepsilon,H}\right|^p\right]=0.
$$
Next, using (\ref{E:pre}), we see that for all $q> 1$,
$$
\limsup_{\varepsilon\downarrow 0}\varepsilon^{2H}\log C^{H}\left(\varepsilon,K_T^{\varepsilon,H}\right)\le-\frac{1}{q}I_T(y).
$$
Therefore, the upper large deviation estimate for the call price follows from the previous inequality.

The proof of Corollary \ref{C:call} is thus completed.

We will next turn our attention to the asymptotic behavior of the implied volatility. It is rather standard to use a call price estimate such as in Corollary \ref{C:call} to study the small-noise and the small-maturity behavior of the implied volatility in the mixed regime. Important results, explaining how to characterize the asymptotic behavior of the implied volatility in various regimes, 
knowing the behavior of the log-call, are contained in the paper \cite{GL} of Gao and Lee.
The authors of \cite{GL} use various parametrizations of the log-moneyness and the dimensionless implied volatility
by a parameter $\theta\rightarrow\infty$. In the small-noise case, we use the parameter $\theta=\varepsilon^{-1}$
and the parametrization $\varepsilon\mapsto k_T^{\varepsilon,H}=y\varepsilon^{\frac{1}{2}-H}$, $\varepsilon\mapsto
\sqrt{\varepsilon}\widehat{\sigma}^H(\varepsilon,k_T^{\varepsilon,H})$. In the small-maturity regime, the parameter 
$\theta=T^{-1}$ is used, and the parametrization is as follows: $T\mapsto k_T=yT^{\frac{1}{2}-H}$, $T\mapsto
\sqrt{T}\widehat{\sigma}(T,k_T)$.
\begin{corollary}\label{C:nakonets}
(i)\,Let $H> 0$, and suppose the conditions in Theorem \ref{T:1} hold. Suppose also that the linear growth condition 
in Lemma \ref{L:environ} is satisfied. Then, for every $y\neq 0$, 
$$
\lim_{\varepsilon\downarrow 0}\widehat{\sigma}^H\left(\varepsilon,k_T^{\varepsilon,H}\right)
=\frac{|y|}{\sqrt{2I_T(y)}}.
$$
(ii)\,Let $0< H< 1$, and suppose the conditions in Theorem \ref{T:smt} hold. Suppose also that the process $\widehat{B}$ 
is $H$-self-similar, and the linear growth condition in Lemma \ref{L:environ} is satisfied. Then, for every $y\neq 0$,
$$
\lim_{T\downarrow 0}\widehat{\sigma}\left(T,k_T\right)
=\frac{|y|}{\sqrt{2\widehat{I}_{\widetilde{T}}(y)}}.
$$
\end{corollary}
\begin{remark}\label{R:find}
Part (ii) of Corollary \ref{C:nakonets} is a generalization of Corollary 4.15 in \cite{FZ}. 
\end{remark}

We will only sketch the proof of part (i) of Corollary \ref{C:nakonets}. The proof of part (ii) is similar. 
Let $L_{\varepsilon}=\log\frac{1}{C^{H}\left(\varepsilon,K_T^{\varepsilon,H}\right)}$. Then it follows from
(\ref{E:call3}) that $\frac{k_T^{\varepsilon,H}}{L_{\varepsilon}}\rightarrow 0$ as $\varepsilon\rightarrow 0$. 
Therefore, we can use formula (7.8) in \cite{GL} in the regime considered in Corollary \ref{C:nakonets}. 
It is not hard to see how to derive the formula in part (i) of Corollary \ref{C:nakonets} from formula (7.8) in \cite{GL}.
 \section{Acknowledgements}\label{S:acq}
I thank Christian Bayer, Peter Friz, Blanka Horvath, Benjamin Stemper, and Stefan Gerhold for interesting and valuable discussions.


\bigskip

\begin{thebibliography}{99}

\bibitem{ALV} E. Al\`{o}s, J. A. Le\'{o}n, and J. Vives. On the short-time behavior of the implied volatility for jump-diffusion models with stochastic volatility. \emph{Finance and Stochastics}, 11 (2007), 571-589.

\bibitem{AP} L. B. G. Andersen and V. V. Piterbarg. Moment explosions in stochastic volatility models. 
\emph{Finance and Stochastics}, (2007), 29-50.

\bibitem{BBK} P. Baldi, G. Ben Arous, and G. Kerkyacharian. Large deviations and the Strassen theorem in
H\"{o}lder spaces. \emph{Stochastic Processes and their Applications}, 42 (1992), 171-180.

\bibitem{BFGMS} C. Bayer, P. Friz, P. Gassiat, J. Martin, and B. Stemper. A regularity structure for rough volatility.
Pre-print, available on arXiv:1710.07481v1, 2017.

\bibitem{BFG} C. Bayer, P. K. Friz, and J. Gatheral. Pricing under rough volatility. \emph{Quantitative Finance}, 16 (2016), 887-904.

\bibitem{BFGHS} C. Bayer, P. Friz, A. Gulisashvili, B. Horvath, and B. Stemper. Short-time near-the-money skew in rough fractional volatility models, submitted for publication, available on arXiv:1703.05132v1, 2017.

\bibitem{BKX} E. Bayraktar, C. Kardaras, and H. Xing. Valuation equations for stochastic volatility models.
\emph{SIAM Journal on Financial Mathematics}, 3 (2012), 351-373.

\bibitem{BLP} M. Bennedsen, A. Lunde, and M. S. Pakkanen. Decoupling the short- and long-term behavior of stochastic volatility. Pre-print, available on arXiv:1610.00332v2, 2017.

\bibitem{CKM} P. Cheridito, H. Kawaguchi, M. Maejima. Fractional Ornstein-Uhlenbeck processes. \emph{Electron. J. Probab.}, 8 (2003), 1-14.

\bibitem{CV1} A. Chronopoulou and F. G. Viens. Estimation and pricing under long-memory stochastic volatility.
\emph{Annals of Finance}, 8 (2012), 379-403.

\bibitem{CV2} A. Chronopoulou and F. G. Viens. Stochastic volatility and option pricing with long-memory in discrete and continuous time. \emph{Quantitative Finance}, 12 (2012), 635-649.

\bibitem{CCR} E. Comte, L Coutin, and E. Renault. Affine fractional stochastic volatility models with
application to option pricing. \emph{Annals of Finance}, 8 (2012), 337-378.

\bibitem{CR} E. Comte and E. Renault. Long memory in continuous-time stochastic volatility models, 
\emph{Mathematical Finance}, 8 (1998), 291-323.

\bibitem{CC} E. Cs\'{a}ki and M. Cs\"{o}rg\H{o}. Inequalities for increments of stochastic processes and moduli of continuity. \emph{The Annals of Probability}, 20 (1992), 1031-1052.

\bibitem{D} L. Decreusefond. Regularity properties of some stochastic Volterra integrals with singular kernels.
\emph{Potential Analysis}, 16 (2002), 139-149.

\bibitem{DU} L. Decreusefond and A. S. \"{U}st\"{u}nel. Stochastic analysis of the fractional Brownian motion.
\emph{Potent. Anal.}, 10 (1999), 177-214.

\bibitem{DeS} F. Delbaen and W. Schachermayer. A general version of the fundamental theorem of asset pricing. \emph{Math. Ann.}, 300 (1994), 463-520.

\bibitem{DZ} A. Dembo and O. Zeitouni. \emph{Large Deviations Techniques and Applications}. Springer Science \& Business Media, 2009.

\bibitem{DFJV} J.-D. Deuschel, P. K. Friz, A. Jacquier, and S. Violante. Marginal density expansions for diffusions and stochastic volatility II: Applications. \emph{Communications on Pure and Applied Mathematics}, 67 (2014), 321-350.

\bibitem{DS} J.-D. Deuschel and D. W. Stroock. \emph{Large Deviations}. Academic Press Boston, 1989.

\bibitem{DD} C. Dol\'{e}ans-Dade. On the existence and unicity of solutions of stochastic integral equations.
\emph{Zeitschrift f\"{u}r Wahrscheinlichkeitstheorie und Verwandte Gebiete}, 36 (1976), 93-101.


\bibitem{ER1} O. El Euch and M. Rosenbaum. The characteristic function of rough Heston models. Pre-print, available on arXiv:1609.02108v1, 2016.

\bibitem{ER2} O. El Euch and M. Rosenbaum. Perfect hedging in rough Heston models. Pre-print, available on arXiv:1703.05049v1, 2017.

\bibitem{ER} S. El Rahouli. Financial modeling with Volterra processes and applications to options, interest rates and credit risk. These, Universit\'{e} de Lorraine, Universit\'{e} du
Luxembourg, 2014.

\bibitem{EM} P. Embrechts and M. Maejima. \emph{Selfsimilar Processes}. Princeton University Press, 2002.

\bibitem{FZ} M. Forde and H. Zhang. Asymptotics for rough stochastic volatility models. \emph{SIAM Journal on Financial Mathematics}, 8 (2017), 114-145.


\bibitem{F2} M. Fukasawa. Short-time at-the-money skew and rough fractional volatility. \emph{Quantitative Finance},
17 (2017), 189-198.

\bibitem{GL} K. Gao and R. Lee. Asymptotics of implied volatility of arbitrary order. \emph{Finance and Stochastics},
18 (2014), 349-392.

\bibitem{GaS1} G. Garnier and K. S\o lna. Correction to Black-Scholes formula due to fractional stochastic volatility.
\emph{SIAM J. Financial Math.}, 8 (2017), 560-588.

\bibitem{GaS2} G. Garnier and K. S\o lna. Option pricing under fast-varying and rough stochastic volatility. Pre-print, available on arXiv:1604.00105v2, 2017.

\bibitem{GJR} J. Gatheral, T. Jaisson, and M. Rosenbaum. Volatility is rough. Pre-print, available on arXiv:1410.3394,
2014.

\bibitem{GJRS} H. Guennoun, A. Jacquier, P. Roome, and F. Shi. Asymptotic behavior of the fractional Heston model.
Pre-print, available on arXiv:1411.7653v2, 2017.

\bibitem{GS1} A. Gulisashvili and E. M. Stein. Asymptotic behavior of the distribution of the stock price in models with stochastic volatility: the Hull-White model. \emph{Comptes Rendus de l'Acad\'{e}mie des Sciences de Paris, S\'{e}rie I}, 343 (2006), 519-523.

\bibitem{GS2} A. Gulisashvili and E. M. Stein. Asymptotic behavior of distribution densities in models with stochastic volatility, I. 
\emph{Math. Finance}, 20 (2010), 447-477.

\bibitem{GS3} A. Gulisashvili and E. M. Stein. Asymptotic behavior of the stock price distribution density and implied volatility in stochastic volatility models. \emph{Applied Mathematics and Optimization}, 61 (2010),  287-315.

\bibitem{G} A. Gulisashvili. \emph{Analytically Tractable Stochastic Stock Price Models}. Springer-Verlag Berlin Heidelberg, 2012.

\bibitem{GVZ1} A. Gulisashvili, F. Viens, and X. Zhang. Extreme-strike asymptotics for general Gaussian stochastic volatility models. Accepted for publication in Annals of Finance, available on arXiv:1502.05442v3, 2017.

\bibitem{GVZ2} A. Gulisashvili, F. Viens, and X. Zhang. Small-time asymptotics for Gaussian self-similar stochastic volatility models. 
\emph{Appl. Math. Optim.} (2018). https://doi.org/10.1007/s00245-018-9497-6, 41 p., available on arXiv:1505.05256, 2016.

\bibitem{Ho} D. Hobson. Comparison results for stochastic volatility models via coupling. \emph{Finance \& Stochastics},
14 (2010), 129-152.

\bibitem{H} H. Hult. Approximating some Volterra type stochastic integrals with application to parameter estimation.
\emph{Stochastic Processes and their Applications}, 105 (2003), 1-32.

\bibitem{Hu} H. Hult. \emph{Extremal behavior of regularly varying stochastic processes}. 
Doctoral Dissertation, Royal Institute of Technology, Stockholm 2003.

\bibitem{JLP} E. A. Jaber, M. Larsson, and S. Pulido. Affine Volterra processes. Pre-print, available on 
arXiv:1708.08796v2, 2017.

\bibitem{JMM} A. Jacquier, C. Martini, and A. Muguruza. On VIX futures in the rough Bergomi models. Available on arXiv:1701.04260v1, 2017.

\bibitem{JPS} A. Jacquier, M. S. Pakkanen, H. Stone. Pathwise large deviations for the rough Bergomi model.
Pre-print, available on arXiv:1706.05291, 2017.

\bibitem{J} B. Jourdain. Loss of martingality in asset price models with lognormal stochastic volatility.
\emph{Internat. J. Theoret. Appl. Finance}, 13 (2004), 767-787.

\bibitem{KS} T. Kaarakka and P. Salminen. On fractional Ornstein-Uhlenbeck processes. \emph{Communications on Stochastic Analysis}, 
5 (2011), 121-133.

\bibitem{KR} I. Karatzas and J. Ruf. Distribution of the time to explosion for one-dimensional diffusions.
\emph{Probability Theory and Related Fields}, 164 (2016), 1027-1069.

\bibitem{KaS} I. Karatzas and S. E. Shreve. \emph{Brownain Motion and Stochastic Calculus}, Second Edition, Springer-Verlag, 1991.

\bibitem{KL} F. Klebaner and R. Lipster. When a stochastic exponential is a true martingale. Extension of the Benes method. \emph{Theory Probab. Appl.}, 58 (2014), 38-62.

\bibitem{Ko} A. N. Kolmogorov. Wienersche Spiralen und einige andere interessante Kurven im Hilbertschen Raum.
\emph{Doklady Acad. USSR}, 26 (1940), 115-118.

\bibitem{LR} M. Larsson and J. Ruf. Notes on the stochastic exponential and logarithm. Pre-print, available on arXiv:1702.03573, 2017.

\bibitem{PL} P. L\'{e}vy. Wiener's random functions, and other Laplacian random functions. Proc. Sec. Berkeley Symp. Math. Statist. Probab., Vol II, University of California Press, Berkeley, CA, 1950, pp. 171-186.

\bibitem{L} A. Lewis. \emph{Option Valuation under Stochastic Volatility: With Mathematica Code}, Finance Pr, 2000

\bibitem{L1} M. Lifshits. \emph{Gaussian Random Functions}. Cluver Academic Publishers Dordrecht Boston London, 1995. 2012.

\bibitem{L2} M. Lifshits. \emph{Lectures on Gaussian Processes}. Springer Verlag, 2012.

\bibitem{LS} S. C. Lim and V. M. Sithi. Asymptotic properties of the fractional Brownian motion 
of Riemann-Liouville type. \emph{Physics Letters A}, 206 (1995), 311-317.

\bibitem{LM} P.-L. Lions and M. Musiela. Correlations and bounds for stochastic volatility models.
\emph{Annales de l'Institut Henri Poincar\'{e} (C) Non Linear Analysis}, 24 (2007), 1-16.

\bibitem{MvN} B. Mandelbrot and J. W. van Ness. Fractional Brownian motions, fractional noises and applications.
\emph{SIAM Review}, 10 (1968), 422-437.

\bibitem{MU} A. Mijatovic and M. Urusov. On the martingale property of certain local martingales. 
\emph{Probability Theory and Related Fields}, 152 (2012), 1-30.

\bibitem{MS-S} A. Millett and M. Sanz-Sol\'{e}. Large deviations of rough paths of the fractional Brownian motion.
\emph{Ann. I. H. Poincar\'{e}-PR}, 42 (2006), 245-271.

\bibitem{MG} G. M. Mol\v{c}an and Yu. I. Golosov, Gaussian stationary processes with asymptotic power spectrum.
\emph{Soviet Math. Dokl.}, 10 (1969), 133-137.

\bibitem{MN} L. Mytnik and E. Neuman. Sample path properties of Volterra processes. \emph{Communications on Stochastic Analysis}, 
6 (2012), 359-377.

\bibitem{NVV} I. Norros, E. Valkeila, and J. Virtamo. An elementary approach to a Girsanov formula and other analytical results on fractional Brownain motions. \emph{Bernoulli}, 5 (1999), 571-587.

\bibitem{P} H. Pham. Large deviations in finance. Third SMAI Uuropean Summer School in Financial Mathematics, Paris 23-27 August, 2010.

\bibitem{Pi} J. Picard. Representation formulae for the fractional Brownian motion. 
\emph{S\'{e}minaire de Probabilit\'{e}s,
Springer-Verlag}, XLIII (2011), 3-70.

\bibitem{RY} D. Revuz and M. Yor. \emph{Continuous Matringales and Brownian Motion}. Springer-Verlag Berlin Heidelberg, 1999.

\bibitem{R} J. Ruf. A new proof for the conditions of Novikov and Kazamaki. \emph{Stoch. Process. Appl.} 123 (2013), 404-421.

\bibitem{Sc} L. Scott. Option pricing when the variance changes randomly: theory, estimation, and an application.
\emph{Journal of Financial and Quantitative Analysis}, 22 (1987), 419-438.

\bibitem{S} C. A. Sin. Complications with stochastic volatility models. \emph{Adv. Appl. Prob.}, 30 (1998), 256-268.

\bibitem{SS} E. M. Stein and J. C. Stein. Stock price distributions with stochastic volatility: An analytic approach. 
\emph{The Review of Finantial Studies}, 4 (1991), 727-752.

\bibitem{UZ} A. S \"{U}st\"{u}nel and M. Zakai, \emph{Transformation of Measure on Wiener Space},
Springer-Verlag, 2000.

\bibitem{VMO} R. Vilela Mendez and M. J. Oliveira. A data-reconstructed fractional volatility model.
available on arXiv:math/0602013v2, 2007.


\bibitem{Y} A. M. Yaglom. \emph{Correlation Theory of Stationary and Related Random Functions, Vol. I}, Springer-Verlag New York, 1987. 

\bibitem{Z} X. Zhang. Euler schemes and large deviations for stochastic Volterra equations with singular kernels. \emph{Journal of
Diff. Equations}, 244 (2008), 2226-2250.



\end{thebibliography}
\end{document}